\documentclass[fleqn,usenatbib]{mnras}

\usepackage{newtxtext,newtxmath}

\usepackage[T1]{fontenc}

\DeclareRobustCommand{\VAN}[3]{#2}
\let\VANthebibliography\thebibliography
\def\thebibliography{\DeclareRobustCommand{\VAN}[3]{##3}\VANthebibliography}

\usepackage{graphicx}	
\usepackage{amsmath}	
\usepackage{soul}
\usepackage{xcolor}

\title[Cut-off effects on the p-mode frequency]{Effects of the photospheric cut-off on the p-mode frequency stability}

\author[D.~Y. Kolotkov et al.]{
Dmitrii Y. Kolotkov$^{1, 2} \thanks{E-mail: D.Kolotkov.1@warwick.ac.uk}$,
Anne-Marie Broomhall$^{1}$,
Amir Hasanzadeh$^{1}$
\\
$^{1}$Centre for Fusion, Space and Astrophysics, Physics Department, University of Warwick, Coventry CV4 7AL, UK\\
$^{2}$Engineering Research Institute \lq\lq Ventspils International Radio Astronomy Centre (VIRAC)\rq\rq, Ventspils University of Applied Sciences, Ventspils, LV-3601, Latvia
}

\date{Accepted XXX. Received YYY; in original form ZZZ}

\pubyear{2024}

\begin{document}
\label{firstpage}
\pagerange{\pageref{firstpage}--\pageref{lastpage}}
\maketitle

\begin{abstract}
Sub-photospheric acoustic resonators allow for the formation of standing p-mode oscillations by reflecting acoustic waves with frequencies below the acoustic cut-off frequency. We employ the Klein-Gordon equation with a piecewise acoustic potential to study the characteristic frequencies of intermediate-degree p-modes, modified by the cut-off effect. For a perfectly reflective photosphere, provided by the infinite value of the acoustic cut-off frequency, characteristic discrete frequencies of the trapped p-modes are fully prescribed by the width of the acoustic potential barrier. Finite values of the acoustic cut-off frequency result in the {reduction} of p-mode frequencies, associated with the decrease in the sound speed by the cut-off effect. For example, for a spherical degree of $\ell = 100$, characteristic p-mode frequencies are found to decrease by up to 200\,$\mu$Hz and the effect is more pronounced for higher radial harmonics. The frequency separation between two consecutive radial harmonics is shown to behave non-asymptotically with non-uniform spacing in the radial harmonic number due to the cut-off effect. We also show how the 11-yr variability of the Sun's photospheric magnetic field can result in the p-mode frequency shifts through the link between the acoustic cut-off frequency and the plasma parameter $\beta$. Using this model, we readily reproduce the observed typical amplitudes of the p-mode frequency shift and its phase behaviour relative to other 11-yr solar cycle proxies. The use of the developed model for comparison with observations requires its generalisation for 2D effects, more realistic profiles of the acoustic potential, and broad-band stochastic drivers.
\end{abstract}




\section{Introduction}\label{sec:intro}

Convection inside the Sun excites acoustic oscillations, which can form standing waves, known as p-modes \citep[see][and references therein for a recent review]{2016LRSP...13....2B}. These p-modes can be readily detected in both Doppler velocity and intensity observations of the {Sun's} photosphere. The spatial structure of the oscillations can be described in terms of spherical harmonics, where the harmonic degree, $\ell$, is determined by the total number of node lines observed at the surface. As the oscillations travel inwards towards the core of the Sun, the waves are refracted. The depth they travel to before returning to the surface, known as the lower turning point (LTP), is dependent on $\ell$, such that the LTP of high-$\ell$ modes is closer to the surface than the LTP of low-$\ell$ modes. Once the oscillations reach the surface, they are reflected by the sharp drop in density. The radius at which the modes are reflected is known as the upper turning point (UTP) and is primarily dependent on mode frequency, such that the UTP is at a larger radius for high-frequency modes than low-frequency modes. The oscillations can, therefore, be considered to be trapped within a cavity, the radial extent of which is determined by the LTP and UTP. The properties of the plasma within those cavities determine the frequencies of the oscillations. 

Although the Sun's convection is capable of exciting oscillations across a wide range of frequencies, there is an upper limit to the frequency at which oscillations are reflected back into the interior, known as the observational acoustic cut-off frequency. As a consequence of this, standing p-mode oscillations are only observed at frequencies below the acoustic cut-off frequency (whereas acoustic waves with frequencies above the acoustic cut off are permitted to propagate out into the solar atmosphere). Although in frequency-power spectra of helio- (or astero-) seismic oscillations, only one acoustic cut-off frequency is observed, which distinguishes between the p-mode and pseudomode regimes, the acoustic cut-off frequency $\omega_\mathrm{ac}$ can also be defined as a parameter that varies as a function of radius, $r$,
\begin{equation}\label{eq_cutoff}
\omega_\mathrm{ac}^2=\frac{c_\mathrm{s}^2}{4H^2}\left(1-2\frac{\textrm{d}H}{\textrm{d}r}\right),
\end{equation}
where $H$ is the density scale height and $c_\mathrm{s}$ is the local sound speed. It can be shown that, when using a standard solar model, $\omega_\mathrm{ac}$ increases with increasing $r$ before reaching a maximum just above the photosphere \citep[e.g.][]{2016LRSP...13....2B}. At a particular radius, only modes with $\omega > \omega_\mathrm{ac}$ are propagate and thus the observational acoustic cut-off frequency corresponds to the maximum value of $\omega_\mathrm{ac}$. This variation in acoustic cut-off explains why different frequency p-modes have different UTPs. 

The WKB approximation can be used to show that for p-modes of spherical degree $\ell$ and radial order $n$,
\begin{equation}\label{eq_WKB}
\int\displaylimits_{r_\mathrm{l}}^{r_\mathrm{u}}\frac{\textrm{d}r}{c_\mathrm{s}}\sqrt{1-\frac{\omega_\mathrm{ac}^2}{\omega^2}-\frac{S_\ell^2}{\omega^2}\left(1-\frac{N^2}{\omega^2}\right)}=\pi\left(\frac{n+\phi_\mathrm{p}}{\omega}\right),
\end{equation}
 \citep[][and references therein]{1984ARA&A..22..593D, 1993afd..conf..399G, 2019ApJ...870...41O}, where $S_\ell$ is the Lamb frequency, $N$ is the Brunt-V\"{a}is\"{a}la frequency, and $r_\mathrm{u}$ and $r_\mathrm{l}$ are the radii of the upper and lower turning points, respectively. Furthermore, $\phi_\mathrm{p}$ represents the phase difference due to near surface reflection, such that if the mode cavity behaves linearly near the turning points, one expects $\phi_\mathrm{p}=-0.5$. In the deep solar interior, where $\omega_\mathrm{ac}\ll\omega$, Eq.~(\ref{eq_WKB}) simplifies to Duvall's law \citep{1982Natur.300..242D}, which demonstrated that $\phi_\mathrm{p}$ is a function of frequency. Unlike Duvall's law, Eq.~(\ref{eq_WKB}) acknowledges the dependence of mode frequency on $\omega_\mathrm{ac}$, such that $\omega_\mathrm{ac}$ determines the location of $r_\mathrm{u}$. As an alternative, this paper aims to investigate the p-mode frequencies obtained from solutions to the radial Klein-Gordon equation, using a finite acoustic potential to represent the acoustic cut-off frequency at the photosphere, which then impacts reflectivity of the oscillations. The focus of this paper is the direct impact acoustic cut-off has on mode frequency. Therefore, in order to isolate this effect from the additional ramifications of changes in the upper turning point, we start by fixing the radius at which the modes are reflected by normalising everything in terms of the length of the modes' respective cavities.

Our ability to model oscillations in the near-surface region, where $r_\mathrm{u}$ is located, is severely limited with observable discrepancies between models and observations that primarily stem from the breakdown of adiabaticity and limitations in modelling near-surface convection and its interaction with the modes \citep[see][for a recent summary of the \lq\lq surface term\rq\rq]{2016LRSP...13....2B}. This, understandably, then impacts our ability to understand the perturbation of oscillation frequencies caused by the presence of a near-surface magnetic field.

The frequencies of the p-mode oscillations are not constant in time. It has now been observed for some time that the p-mode frequencies vary systematically through the Sun's 11-yr Schw\"{a}be magnetic activity cycle \citep[e.g.][]{1985Natur.318..449W, 1989A&A...224..253P, 1990Natur.345..322E, 1990Natur.345..779L}. The frequency variation has an amplitude of around $0.5\,\rm\mu Hz$ and is in phase with magnetic activity. However, the exact magnitude of the change in frequency is dependent on both $\ell$ and the frequency of the mode \citep[see][and references therein]{2014SSRv..186..191B}. Interestingly, \citet{2011ApJ...743...99J} found that the acoustic cut-off frequency {itself} varies in phase with the solar cycle. 

Broadly speaking, the influence of the magnetic fields on the oscillations frequencies is often attributed to the direct impact of the Lorentz force on the modes and the indirect effect of magnetic fields on the properties of the acoustic cavities and the plasma contained within them. However, the exact mechanism responsible for this variation is still not well understood. A number of papers suggest that the direct impact of magnetic fields on the oscillations is too small to alone explain the observed variation \citep[e.g.][]{2005ApJ...625..548D, 2005A&A...439..713F}. However, \citet{2018ApJ...854...74K} demonstrated that, unless the near-surface magnetic field is very strong (such that the plasma $\beta<1$), the indirect perturbation on oscillation frequencies is far smaller than the perturbation due to the direct effect. Furthermore, the authors find that, for all magnetic field configurations included in their study, the indirect effect acts to decrease the oscillation frequencies i.e. changes them in the opposite sense to the variations observed through the solar cycle. However, \citet{2018ApJ...854...74K} were able to approximately replicate the magnitude and frequency dependence of the frequency shifts by invoking a $10\,\rm kG$ decrease in toroidal magnetic field from cycle minimum to maximum. In this paper, we consider the effect of a magnetic field on the acoustic cut-off frequency and determine consequences of a variation in the acoustic cut-off frequency on the modelled p-mode frequencies.

Section \ref{sec:model} describes the model used in this paper and the main assumptions made. Section \ref{Sec:trapped} describes the impact of varying the acoustic cut-off on the frequencies of trapped p-modes, while Section \ref{Sec:shifts} demonstrates that varying the acoustic cut-off in a physically motivated way can replicate properties of observed solar cycle frequency shifts, such as the magnitude and in-phase variation. Finally, Section \ref{sec:disc} gives the main conclusions of this work.

\section{Model and assumptions}
\label{sec:model}

\begin{figure}
    \centering
    \includegraphics[width=\columnwidth]{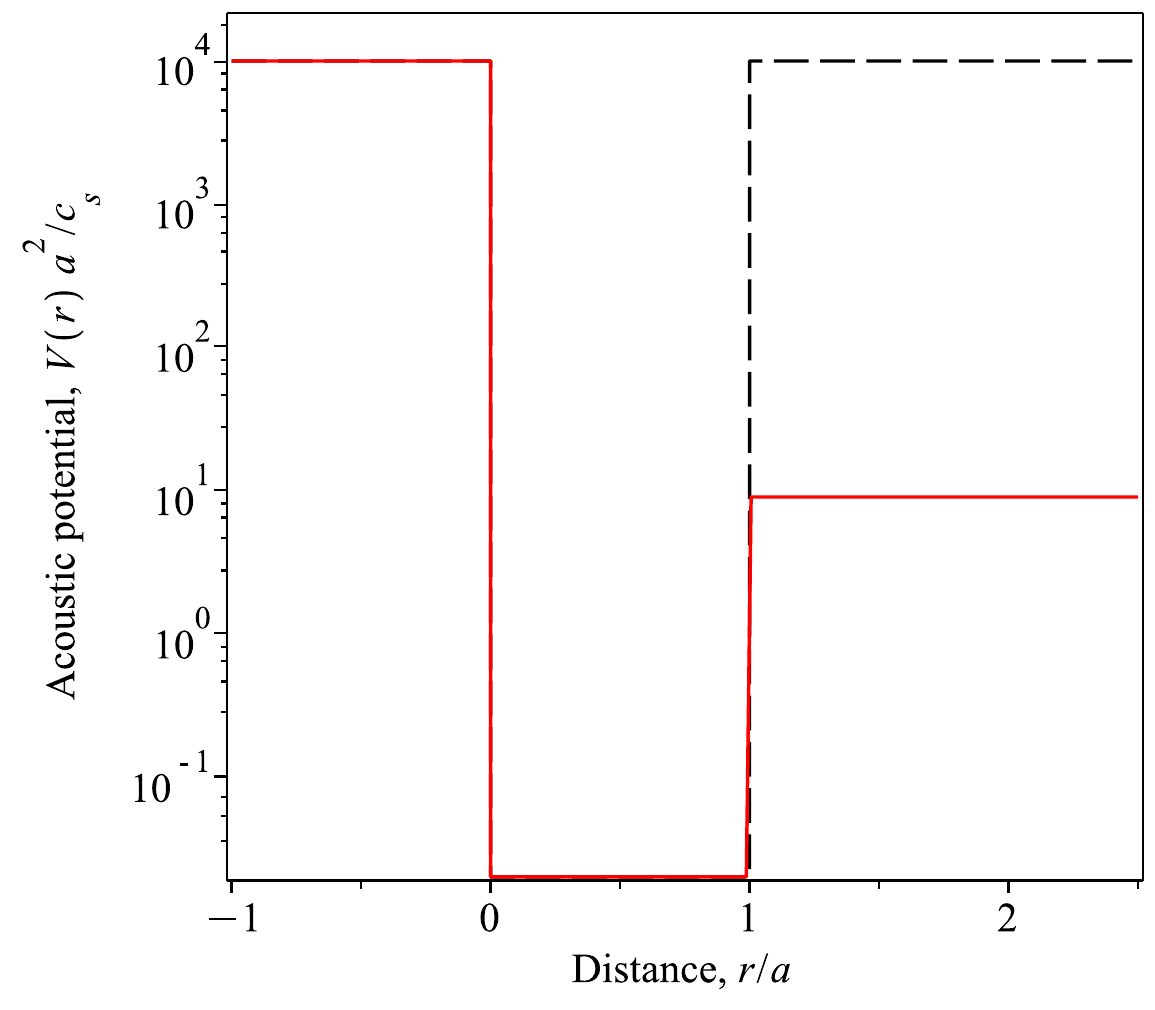}
    \caption{Acoustic potential $V(r)$ given by Eq.~(\ref{eq:potential}), with partial (red) and infinite (black dashed) reflectivity at the photosphere, $r=a$. The width $a$ of the potential $V(r)$ is determined by the effective penetration depth of acoustic waves.}
    \label{fig:potential}
\end{figure}

The dynamics of acoustic waves with spherical degrees $\ell \lesssim 100$ in a convection zone cavity can be considered predominantly vertical and described by the following Klein-Gordon equation \citep[see e.g.][and references therein]{1991ApJ...375L..35K, 1995MNRAS.272..850R, 1998MNRAS.298..464V, 1998ApJ...505L..51N, 2008SoPh..251..523T},
\begin{equation}\label{eq:KG}
    \frac{\partial ^2\psi}{\partial r^2}-\frac{\partial ^2\psi}{\partial t^2} - V(r)\psi = f(t,r),
\end{equation}
where $\psi(t,r)$ is a particular spherical component of the wave-caused perturbation, $V(r)$ is the potential barrier which represents an effective subphotospheric acoustic resonator, and $f(t,r)$ is the source function which, in general, mimics the excitation of acoustic waves by convective zone motions. Non-adiabatic effects responsible for the wave damping are omitted. In Eq.~(\ref{eq:KG}), we normalise the sound speed $c_\mathrm{s}$ and the acoustic cavity depth $a$ to unity, so that the characteristic wave frequencies described by Eq.~(\ref{eq:KG}) are measured in units of the inverse acoustic transit time $1/\tau_\mathrm{A}$\footnote{Estimated, for example, through the standard solar model $c_\mathrm{s}^2\approx (\gamma - 1)GM_\odot(r^{-1}-R_\odot^{-1})$ \citep[e.g.][Eq.~(43)]{2021LRSP...18....2C} as $\tau_\mathrm{A}=\int_{R_\odot-a}^{R_\odot}dr/c_\mathrm{s}$, where $R_\odot-a$ is the lower turning point. For example, for $\ell = 100$ and oscillation frequency around 3000\,$\mu$Hz (radial harmonic $n=7$), $a\approx 0.1 R_\odot$ \citep[e.g.][Figs.~3.14 and 7.13]{2010aste.book.....A} which gives $\tau_\mathrm{A}\approx 20$\,min.}. For the acoustic potential $V(r)$, we choose a piecewise form,
\begin{equation}\label{eq:potential}
    V(r) = \begin{cases}
			 \alpha^2, & \text{for $r \ge a$ }\\
             0, & \text{for $0<r<a$} \\
            \gg \alpha^2, & \text{for $r\le 0$}
		 \end{cases}
\end{equation}
which is illustrated in Fig.~\ref{fig:potential}. Here, the parameter $\alpha$ represents the effective acoustic cut-off frequency at the photospheric level, $\alpha = \omega_\mathrm{ac}(r=a)$ \citep[cf. Fig. 5 in][for example]{2012ApJ...758...43B}. Such a form of $V(r)$ implies an acoustic resonator of a Fabry-P\'erot type, i.e. waves of practically all frequencies reflect from the lower boundary of the  resonator, $r=0$, while at the photospheric level, $r=a$, waves with frequencies below/above the acoustic cut-off $\alpha$ reflect/propagate, respectively.
In this piecewise form of $V(r)$, we neglect the variation of the acoustic cut-off frequency $\alpha$ with height, so that all waves with frequencies below $\alpha$ reflect at the same photospheric level, $r=a$.
We also stress that the lower boundary of $V(r)$ at $r=0$ only mimics the lower turning of p-modes, as the described model is intrinsically one-dimensional and, hence, cannot properly address the effect of wave refraction.

\begin{figure*}
    \centering
    \includegraphics[width=0.32\textwidth]{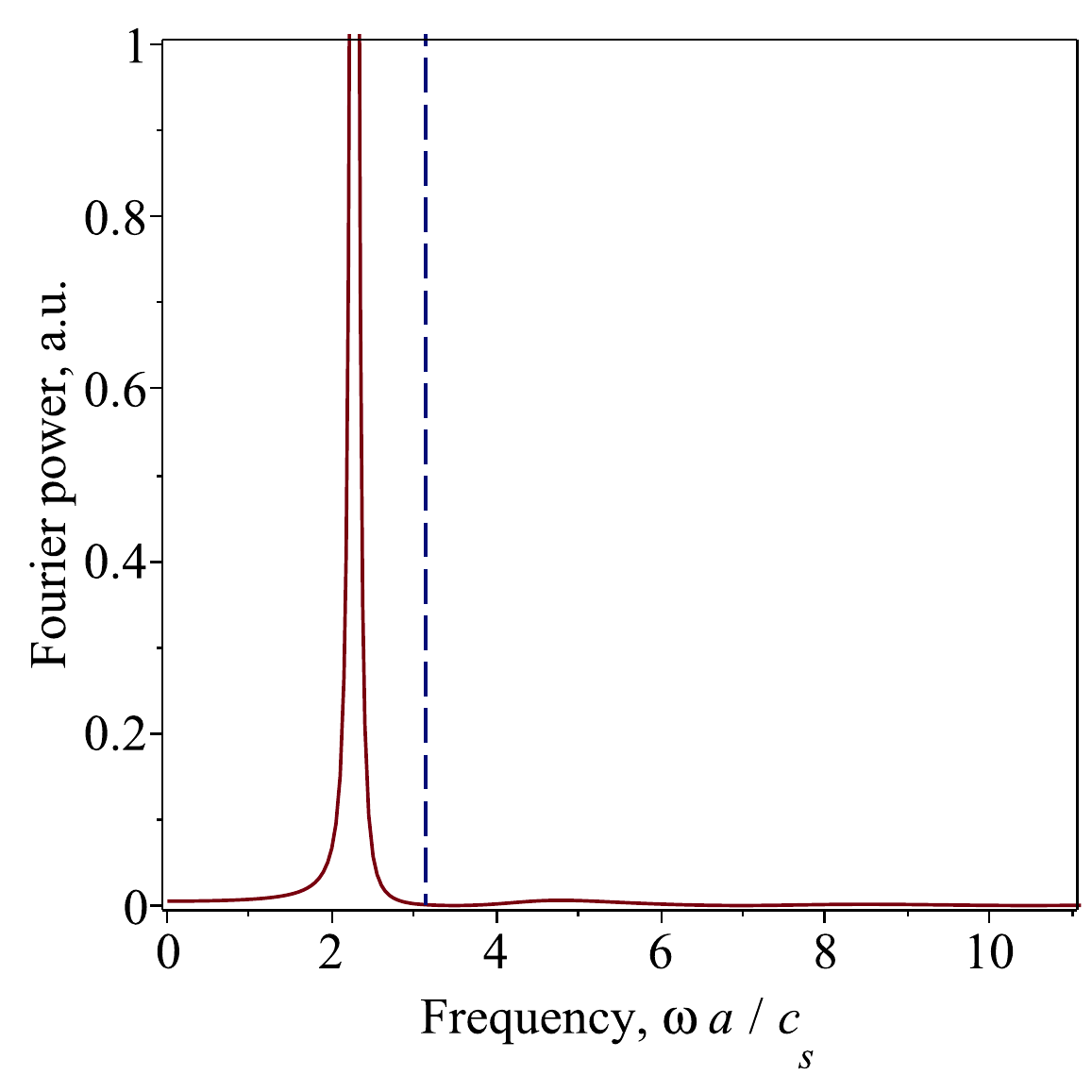}
    \includegraphics[width=0.32\textwidth]{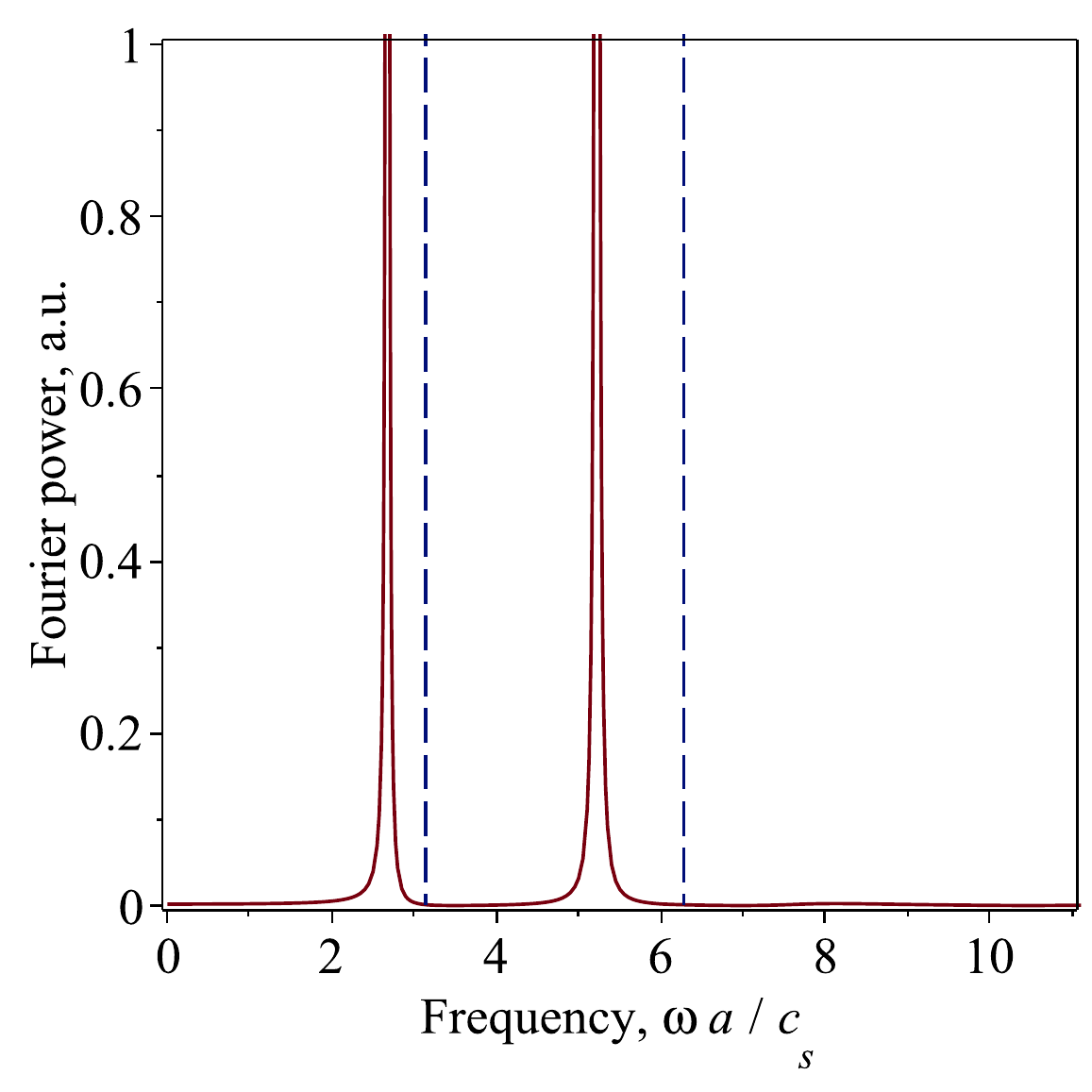}
    \includegraphics[width=0.32\textwidth]{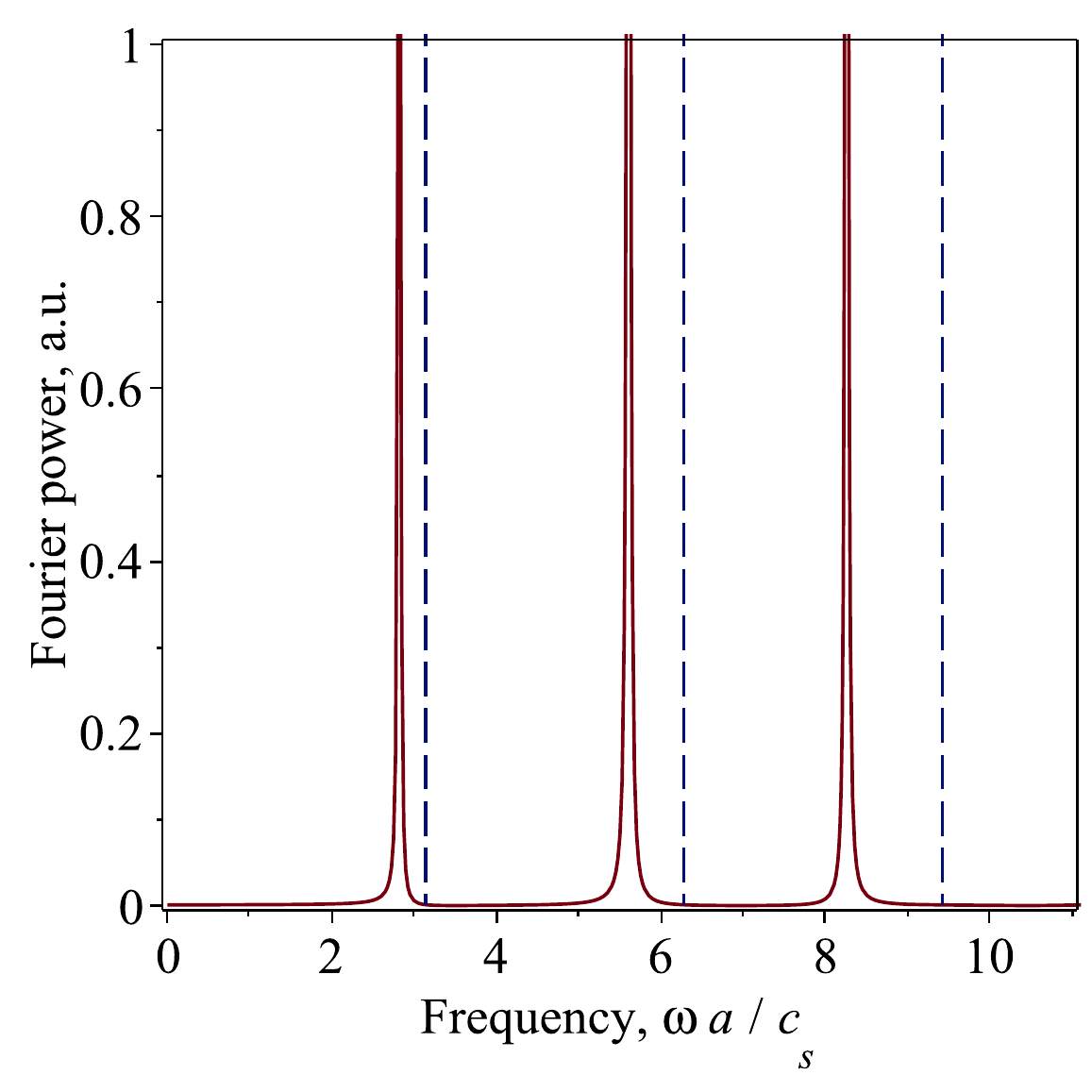}
    \caption{Fourier power spectra given by Eq.~(\ref{eq:spectrum}) for acoustic waves trapped within the acoustic cavity illustrated in Fig.~\ref{fig:potential}, with $r_0=0.9a$, $f_0=1$\,a.u. (the addressed problem is linear), and the acoustic cut-off angular frequency $\alpha =$3 (left), 6 (middle), and 9 (right) normalised to the acoustic travel time across the cavity, $\tau_\mathrm{A}\approx 20$\,min (see Sec.~\ref{sec:model}). The vertical dashed lines in each panel indicate the values of trapped mode frequencies, $n\pi/\tau_\mathrm{A}$ in the absence of the cut-off effect.}
    \label{fig:spectra}
\end{figure*}

Using the exact analytical solution to Eq.~(\ref{eq:KG}) given in e.g. \citet{1991ApJ...375L..35K} for a harmonic point source $f(t,r)$, the Fourier power spectrum of the wave function $\psi(t,r)$ in the vicinity of $r=a$ can be written as,
\begin{equation}\label{eq:spectrum}
    \mathcal{F}(\omega) = \left| \frac{f_0 \sin{\omega r_0}}{\omega\cos{\omega}+\sqrt{\alpha^2 - \omega^2}\sin{\omega}} \right|^2,
\end{equation}
where $f_0$ and $r_0$ are the source's amplitude and position, and $\omega$ is the angular frequency of the excited acoustic wave. The typical shape of the Fourier power spectrum prescribed by Eq.~(\ref{eq:spectrum}) is shown in Fig.~\ref{fig:spectra} for $\alpha = 3$, 6, and 9 of the inverse acoustic transit time $1/\tau_\mathrm{A}$, with one, two, and three standing radial harmonics with $\omega<\alpha$ excited within the cavity ($0<r<a$), respectively. Outside the cavity ($r>a$) these modes evanesce with the characteristic $e$-folding length $\lambda_e=c_\mathrm{s}/\sqrt{\alpha^2 - \omega^2}$. For example, for the p-mode frequency $\omega/2\pi = 3000$\,$\mu$Hz, cut-off frequency $\alpha/2\pi = 5000$\,$\mu$Hz, and sound speed at the photosphere $c_\mathrm{s}=10$\,km/s, we obtain $\lambda_e\approx 400$\,km, i.e. the perturbation caused by the p-mode trapped in the acoustic cavity below the photosphere can reach about the temperature minimum region in the solar atmosphere. This is in broad agreement with observations, where the amplitude of oscillations observed in different intensity measurements at different wavelengths decreases if observations are made higher in the solar atmosphere. For example, Solar and Heliospheric Observatory's (SoHO's) Variability of Solar IRradiance and Gravity
Oscillations (ViRGO) made unresolved intensity observations of the Sun at three different wavelengths. The response functions of these observations peak at heights, separated by a few 10s of km \citep{2005ApJ...623.1215J}. In turn, the oscillation amplitudes drop by approximately a factor of $e$ \citep{1997SoPh..170....1F}.

In this work, we consider the acoustic cut-off frequency $\alpha$ as a free parameter characterising the partial reflectivity of the photosphere, and investigate the associated frequency corrections and modulation of resonant acoustic waves trapped inside the cavity as effective p-modes.
For $\omega>\alpha$, the propagating acoustic waves can also form peaks in the Fourier power spectrum due to the effect of constructive interference, which is known as the phenomenon of pseudo-modes \citep[see e.g.][and references therein]{2022MNRAS.512.5743K}. However, the further discussion of pseudo-modes is out of the scope of this work.

\section{P-mode frequency corrections by acoustic cut-off}
\label{Sec:trapped}

Using Eq.~(\ref{eq:spectrum}), the peaks in the lower-frequency ($\omega<\alpha$) part of the Fourier spectrum shown in Fig.~\ref{fig:spectra}, occur when the denominator on its right-hand side goes to zero. Thus, the eigenfrequencies of trapped p-modes in the acoustic resonator given by Eq.~(\ref{eq:potential}) can be determined from the following transcendental algebraic equation,
\begin{equation}\label{eq:disp}
    \tan{\omega} + \frac{\omega}{\sqrt{\alpha^2-\omega^2}}=0.
\end{equation}
As such, the p-mode eigenfrequencies seem to be generally independent of the source position $r_0$ except for cases with $r_0 = a/n$ where $n$ is the integer number, resulting in the preferential excitation of even or odd radial harmonics. Thus, the acoustic cut-off frequency $\alpha$ appears as the only free parameter in Eq.~(\ref{eq:disp}) (given that the sound speed $c_\mathrm{s}$ and the acoustic cavity depth $a$ {are set to unity} for normalisation). Indeed, for a perfectly reflecting photosphere with $\alpha \to \infty$, the p-mode eigenfrequencies are determined {by the condition $\tan{\omega} = 0$}, resulting in $ \omega = n\pi/\tau_\mathrm{A}$, i.e. are fully prescribed by the acoustic cavity depth $a$ and the sound speed $c_\mathrm{s}$. Finite values of $\alpha$, in turn, would lead to a modification in the characteristic frequencies of the fundamental and higher radial p-mode harmonics, as seen, for example, in Fig.~\ref{fig:spectra}.

\begin{figure*}
    \centering
    \includegraphics[width=0.48\textwidth]{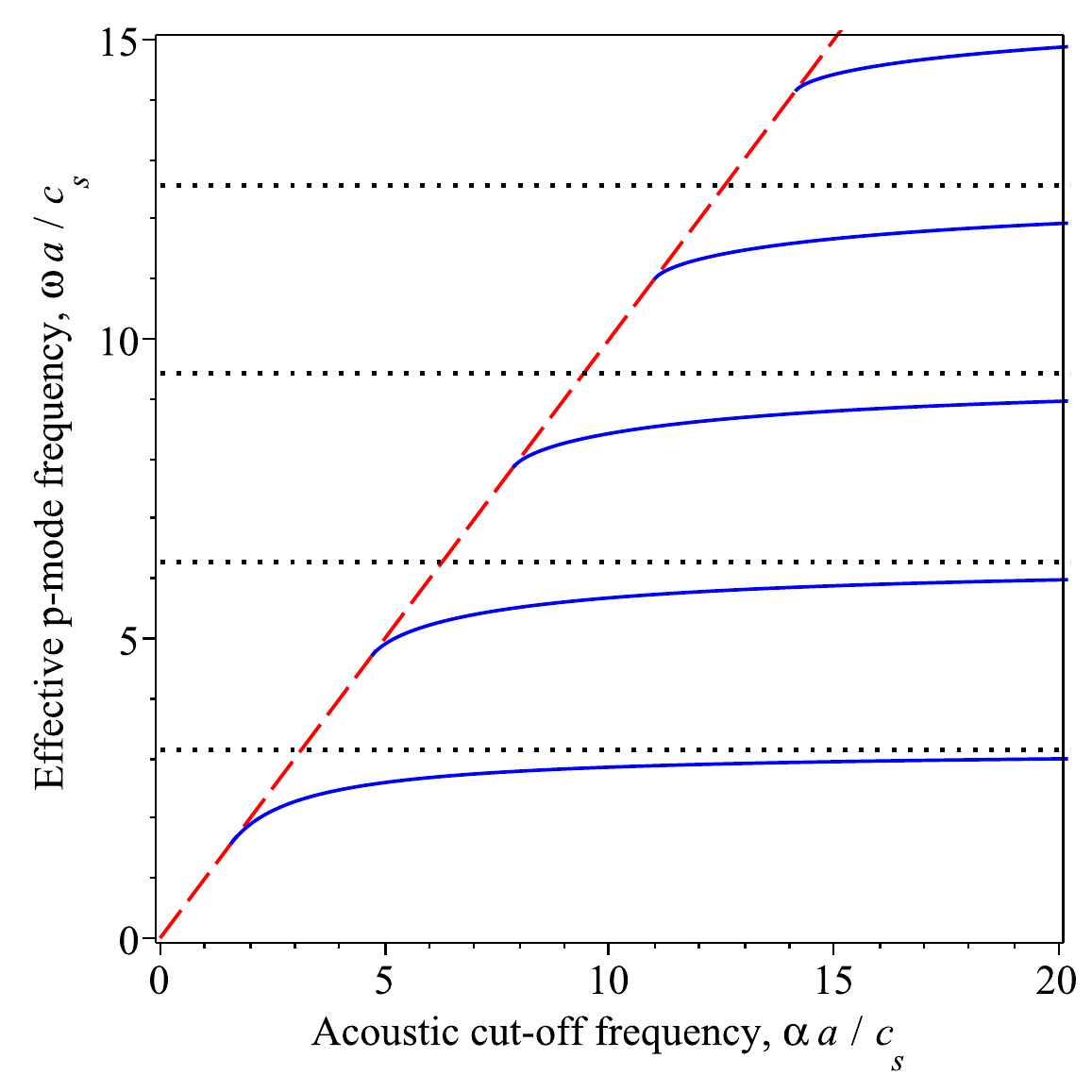}
    \includegraphics[width=0.48\textwidth]{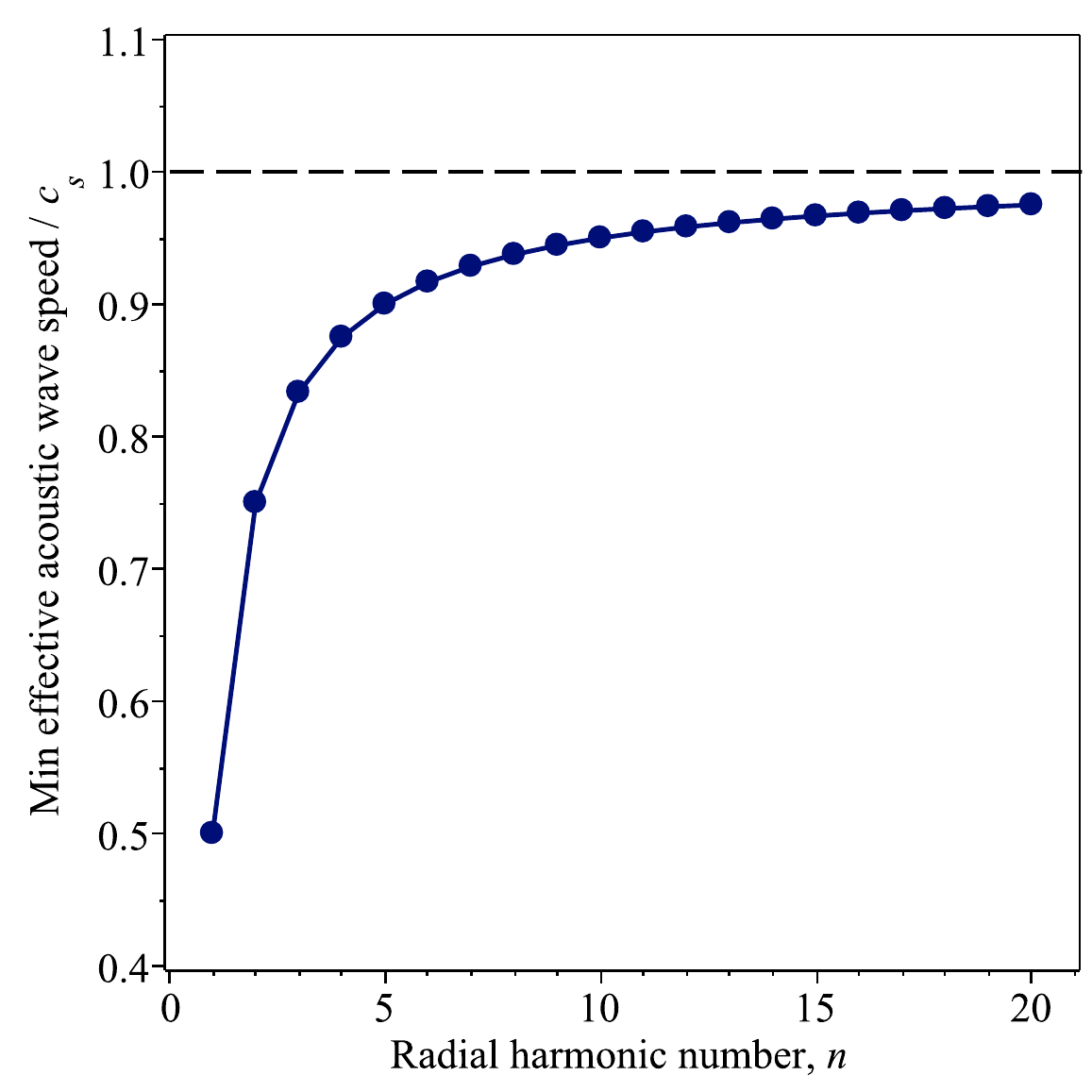}
    \caption{Left: Normalised angular frequency (blue) of the first five ($n=1$--5) radial harmonics of acoustic waves trapped within the acoustic cavity (Fig.~\ref{fig:potential}) vs. normalised acoustic cut-off angular frequency, obtained with Eq.~(\ref{eq:disp}). The horizontal dotted lines indicate the trapped mode frequencies ($\propto n\pi$) in the absence of the cut-off effect {(as $\alpha$ tends to infinity)}. The red dashed line shows $\omega=\alpha$.
    Right: Decrease in the effective speed of trapped acoustic waves vs. radial harmonic number, caused by the cut-off effect. The minimum effective acoustic wave speed is determined as $(\pi/2 + [n-1]\pi)/n\pi$ (normalised to $c_\mathrm{s}$, see Sec.~\ref{Sec:trapped}).}
    \label{fig:p-mode_freq}
\end{figure*}

To assess the modification of the p-mode frequencies caused by the effect of finite reflectivity of the photosphere (finite $\alpha$), we treat Eq.~(\ref{eq:disp}) numerically in the mathematical software package \textit{Maple 2022} and visualise solutions for the first five radial harmonics in Fig.~\ref{fig:p-mode_freq} (left panel). For some fixed finite value of the acoustic cut-off $\alpha$, the obtained effective p-mode frequencies are clearly seen to deviate from $n\pi$ (normalised to $1/\tau_\mathrm{A}$). Furthermore, the magnitude of the deviation increases with the radial harmonic number, $n$.
{ More specifically, when $\omega=\alpha$ (see where the blue and red lines meet in the left-hand panel of Fig.~\ref{fig:p-mode_freq}), the effective p-mode frequency equals $\pi/2 + (n-1)\pi$. As $\alpha$ then increases, the p-mode frequency, for each radial harmonic, tends to $n\pi$.}
The latter means, in particular, that for each radial harmonic, the maximum deviation of the p-mode {angular} frequency from $n\pi$ is $\pi/2$ {(normalised to $1/\tau_\mathrm{A}$)} which is independent of the harmonic number, $n$. {For $\tau_\mathrm{A}=20$\,min (see Sec.~\ref{sec:model}), this results in the maximum decrease in the p-mode cyclic frequency about 200\,$\mu$Hz, obtained as $\pi/2/2\pi/\tau_\mathrm{A}$.}
{The closer the acoustic cut-off frequency is to $n\pi$, the bigger the deviation.}
The physical nature of this effect is connected with the decrease of the effective acoustic wave speed with $\alpha$. Indeed, the effective acoustic wave speed can be determined as $\omega_n/k_n$ where $k_n$ is the wavenumber of the $n$-th radial harmonic. For trapped p-modes, $k_n$ is fully prescribed by the cavity size $a$ as $k_n = n\pi/a$ and thus is independent of the acoustic cut-off frequency, $\alpha$. The right-hand panel in Fig.~\ref{fig:p-mode_freq} shows the minimum possible effective acoustic wave speed, determined as $(\pi/2 + [n-1]\pi)/n\pi$ (normalised to $c_\mathrm{s}$), for the first twenty radial harmonics. According to it, the effective acoustic wave speed can differ from the standard sound speed $c_\mathrm{s}$ by 20\%--50\% for lower radial harmonics and by just a few per cent for higher radial harmonics due to the effect of $\alpha$. For all $n$s, this modification in the effective acoustic wave speed with $\alpha$ results in up to $\pi/2$ modification in the effective p-mode frequency, as pointed out above.

\begin{figure*}
    \centering    
    \includegraphics[width=0.48\textwidth]{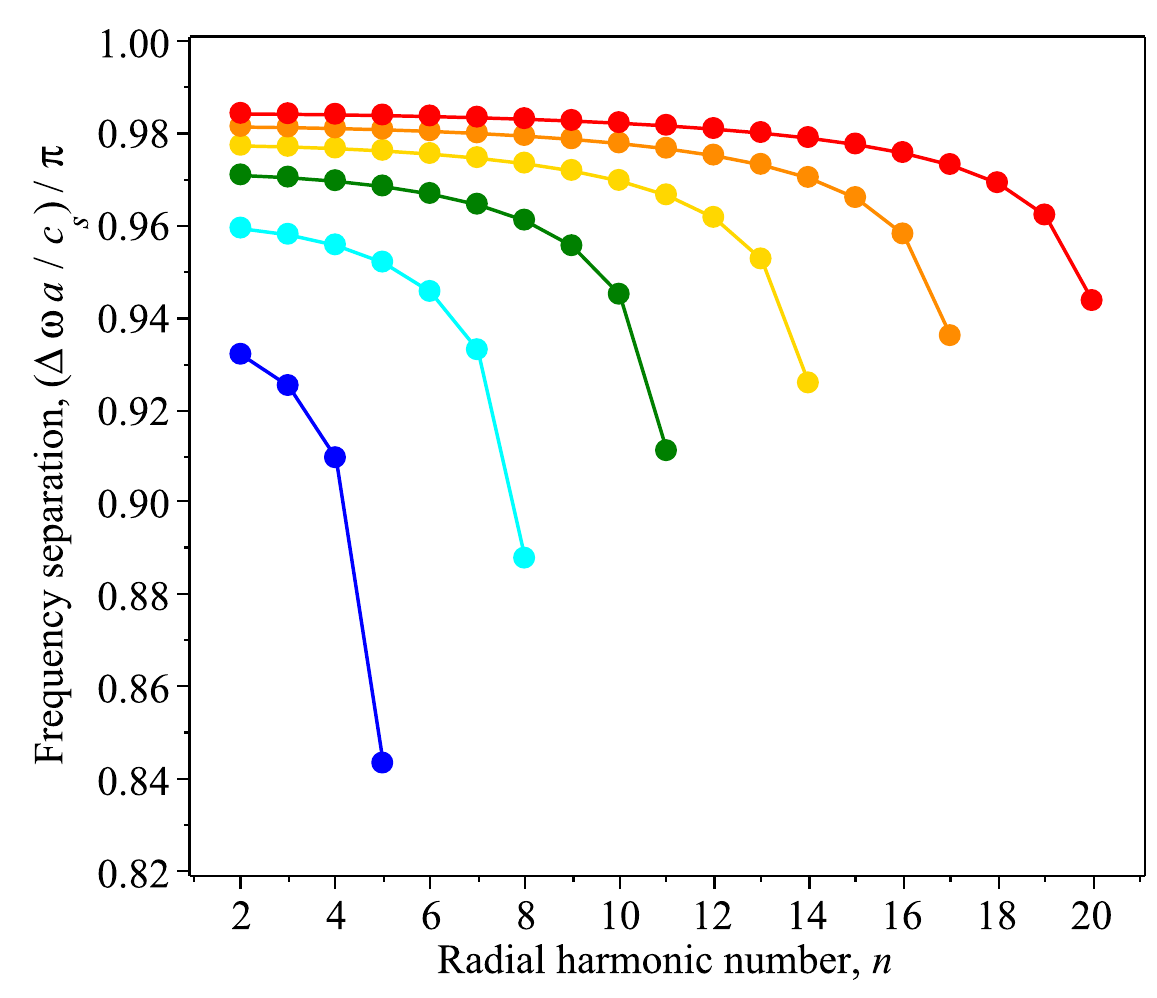}
    \includegraphics[width=0.48\textwidth]{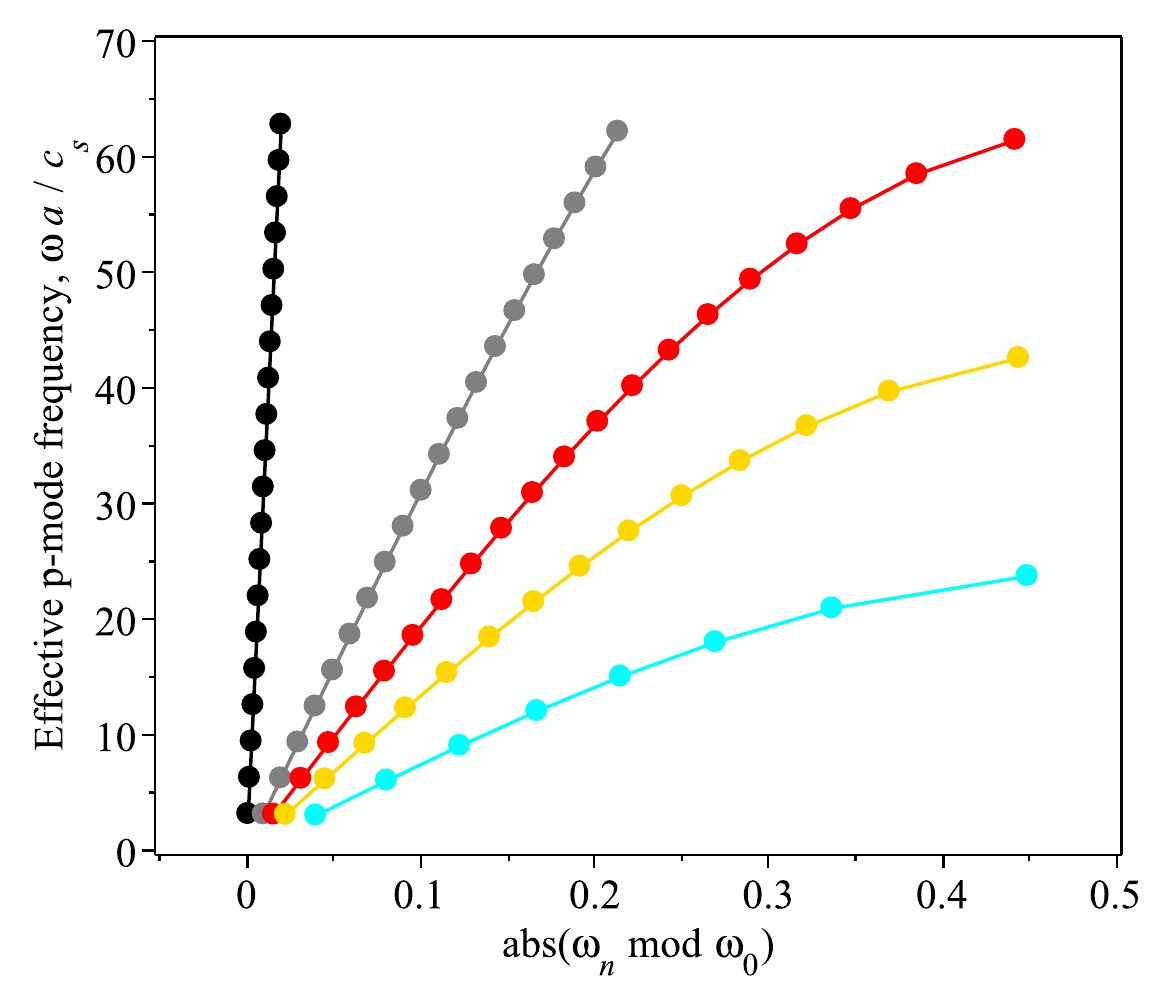}
    \caption{Left: Frequency separation between modes of successive radial harmonics ($\Delta \omega_n = \omega_n - \omega_{n-1}$, also known as large frequency separation) shown in the left panel of Fig.~\ref{fig:p-mode_freq} vs. radial harmonic number, for the normalised acoustic cut-off angular frequency $\alpha\tau_\mathrm{A}=14.4$ (blue, $n=1$--5 excited), 24 (cyan, $n=1$--8 excited), 33.6 (green, $n=1$--11 excited), 43.3 (yellow, $n=1$--14 excited), 52.9 (dark orange, $n=1$--17 excited), and 62.5 (red, $n=1$--20 excited).
    Right: Echelle diagram (for the first several radial harmonics excited) illustrating the dependence of the effective p-mode frequency $\omega_n$ on the corresponding frequency correction $|\varepsilon_n|$ caused by the cut-off effect, obtained as modulo between $\omega_n$ and $\omega_0$ (see Eq.~(\ref{eq:epsilon_n})), for $\alpha\tau_\mathrm{A}=1000$ (black, effect is practically absent), $\alpha\tau_\mathrm{A}=100$ (grey), $\alpha\tau_\mathrm{A}=62.5$ (red), $\alpha\tau_\mathrm{A}=43.3$ (yellow), and $\alpha\tau_\mathrm{A}=24$ (cyan).}
    \label{fig:p-mode_separation}
\end{figure*}

\begin{figure}
    \centering    
    \includegraphics[width=\columnwidth]{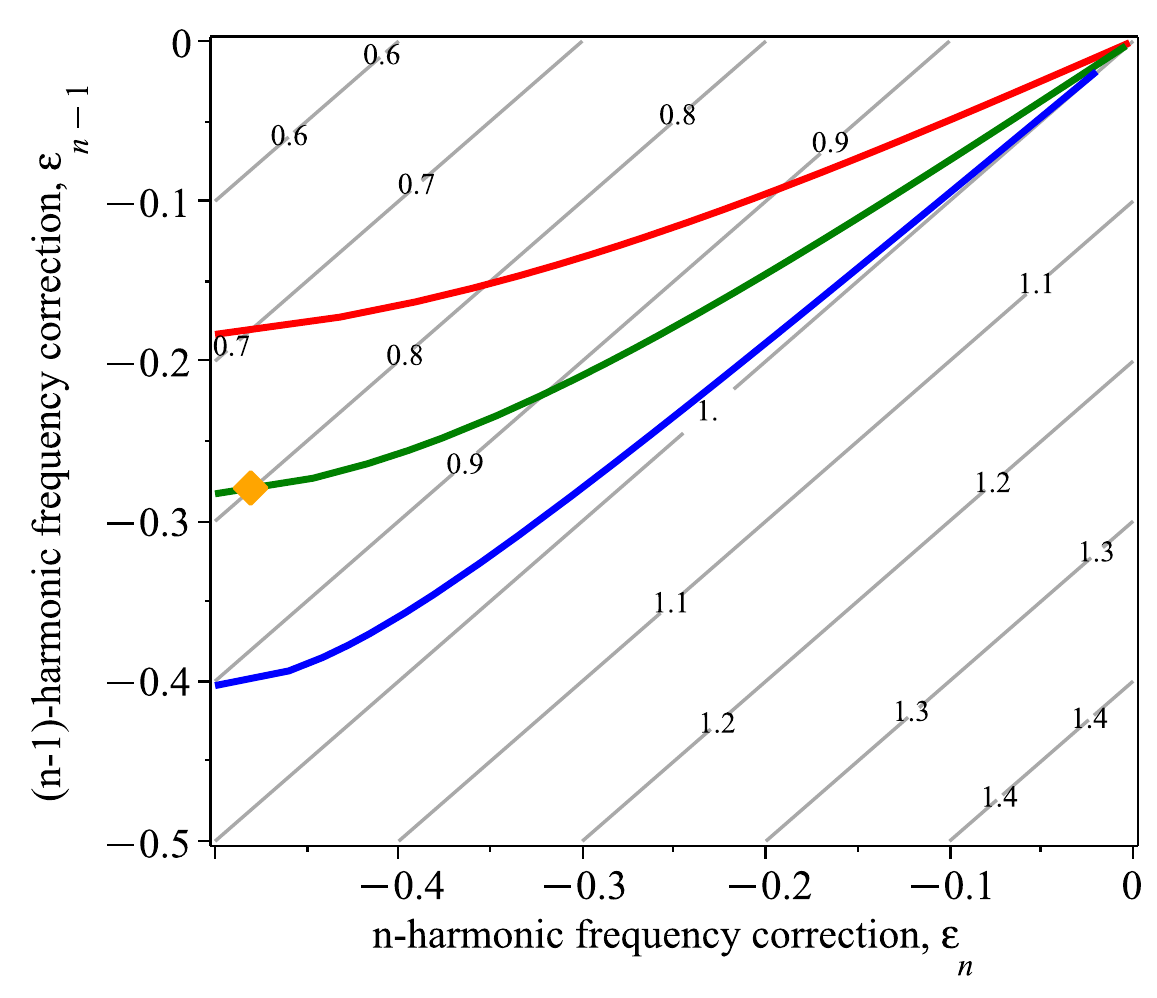}
    \caption{Contour-plot (thin grey lines) of the parameter $\Delta \omega_n/\omega_0$ characterising the possible departure of the frequency separation $\Delta \omega_n$ from $\omega_0 = \pi /\tau_\mathrm{A}$ (the value of $\Delta \omega_n$ in the absence of the cut-off effect), caused by the frequency corrections $\varepsilon_n$ and $\varepsilon_{n-1}$ introduced in Eqs.~(\ref{eq:epsilon_n})--(\ref{eq:epsilon_n-1}). The red, green, and blue thick lines show the parametric curves of $\varepsilon_n$ and $\varepsilon_{n-1}$ caused by the cut-off effect for pairs of successive radial harmonics $n=$ 2 and 1 (red), 4 and 3 (green), 20 and 19 (blue) and the normalised acoustic cut-off angular frequency $\alpha\tau_\mathrm{A}$ varying respectively from 4.7 (red), 11 (green), and 61.3 (blue) to 1000 (all). For $\alpha\tau_\mathrm{A}\to\infty$, the red, green, and blue curves tend to the $\Delta \omega_n/\omega_0 = 1$ contour. The yellow diamond shows $\varepsilon_4$ and $\varepsilon_3$ for $\Delta \omega_n/\omega_0 = 0.8$, for illustration.}
    \label{fig:p-mode_sep_corr}
\end{figure}

We now consider the impact of $\alpha$ on the frequency separation $\Delta \omega_n = \omega_{n} - \omega_{n-1}$ between two consecutive radial harmonics, referred to as the \textit{large frequency separation} in observations {\citep[see][and references therein for a recent review in the context of asteroseismology]{2019LRSP...16....4G}.} The left-hand panel in Fig.~\ref{fig:p-mode_separation} shows $\Delta \omega_n$ estimated from the numerical solution of Eq.~(\ref{eq:disp}) (see Fig.~\ref{fig:p-mode_freq}, left panel) for several values of the cut-off frequency $\alpha$. First, we observe that $\Delta \omega_n$ predicted by Eq.~(\ref{eq:disp}) appears to be smaller than $\pi$ (in normalised units) for all cases considered. Also, $\Delta \omega_n$ is found to be generally non-uniform, i.e. it {decreases} with the radial harmonic number $n$. On the other hand, this non-uniformity seems to be less pronounced for higher $\alpha$, when more radial harmonics are excited.
For example, for $\alpha \tau_\mathrm{A} \approx 14.4$ only five trapped radial harmonics exist and $\Delta \omega_n$ decreases by about 10\% for $n$ varying from 1--2 to 4--5. In contrast, for $\alpha \tau_\mathrm{A} \approx 62.5$, this decrease of $\Delta \omega_n$ with $n$ (from several lowest-$n$ harmonics up to $n=20$) is found to be about 4\%.

For practical purposes, such behaviour of $\Delta \omega_n$ with the radial harmonic number $n$ and the cut-off frequency $\alpha$ motivates us to represent $\omega_n$ and $\omega_{n-1}$ through the constant frequency $\omega_0 = \pi /\tau_\mathrm{A}$ (prescribed by the acoustic travel time $\tau_\mathrm{A}$ through the cavity) and some dimensionless {non-asymptotic corrections} $\varepsilon_n(\alpha)$ and $\varepsilon_{n-1}(\alpha)$ as
\begin{align}
&\omega_n = (n + \varepsilon_n)\omega_0,\label{eq:epsilon_n}\\
&\omega_{n-1} = (n - 1 + \varepsilon_{n-1})\omega_0,\label{eq:epsilon_n-1}
\end{align}
which is convenient for determining frequency corrections $\varepsilon_n$ and $\varepsilon_{n-1}$ caused by the effect of acoustic cut-off from observations.
{In contrast to the asymptotic regime \citep[e.g.][]{2013A&A...550A.126M}, $\varepsilon_n$ and $\varepsilon_{n-1}$ may in general vary with both $n$ and $\alpha$ in Eqs.~(\ref{eq:epsilon_n})--(\ref{eq:epsilon_n-1}).}
The right-hand panel of Fig.~\ref{fig:p-mode_separation} shows how the cut-off effect is pronounced in the standard echelle diagram, where the departure from the vertical shape increases with the decrease in $\alpha$.
Furthermore, one can consider a potentially observable dimensionless parameter $\Delta \omega_n/\omega_0 = 1 + \varepsilon_n - \varepsilon_{n-1}$ and assess its expected values treating $\varepsilon_n$ and $\varepsilon_{n-1}$ as free parameters (see grey contour lines in Fig.~\ref{fig:p-mode_sep_corr}). Thus, $\Delta \omega_n/\omega_0$ is expected to equal unity for $\varepsilon_n = \varepsilon_{n-1}$; to be $>1$ for $\varepsilon_n > \varepsilon_{n-1}$; and to be $<1$ for $\varepsilon_n < \varepsilon_{n-1}$. Substituting our numerical solution for $\omega_n$ and $\omega_{n-1}$ shown in Fig.~\ref{fig:p-mode_freq} (left panel) to Eqs.~(\ref{eq:epsilon_n})--(\ref{eq:epsilon_n-1}), we obtain that $-0.5\le \varepsilon_n < \varepsilon_{n-1}<0$ for finite $\alpha$ and $\varepsilon_n \to \varepsilon_{n-1} \to 0$ for $\alpha \to \infty$, resulting in $\Delta \omega_n/\omega_0 \lesssim 1$. For illustration, Fig.~\ref{fig:p-mode_sep_corr} shows $\varepsilon_n$ vs. $\varepsilon_{n-1}$ caused by the effect of acoustic cut-off for the pairs of radial harmonics with $n=2$ and 1, $n=4$ and 3, $n=20$ and 19. The intersection of those lines with the grey contours of the parameter $\Delta \omega_n/\omega_0$ provides the corresponding values of the frequency correction parameters $\varepsilon_n$ and  $\varepsilon_{n-1}$.
Thus, if one has the value of $\Delta \omega_n/\omega_0$ for a given pair of radial harmonics $n$ and $n-1$ (and fixed spherical degree $\ell$) from observations, Fig.~\ref{fig:p-mode_sep_corr} can potentially be used for constraining the frequency corrections $\varepsilon_n$ and  $\varepsilon_{n-1}$, caused by the effect of the photospheric acoustic cut-off\footnote{For example, for $\omega_0/2\pi \approx 416$\,$\mu$Hz ($\tau_\mathrm{A}=20$\,min) and the large separation $\Delta \omega_n/2\pi = 330$\,$\mu$Hz between radial harmonics $n=4$ and 3 with fixed spherical degree $\ell=100$, we obtain $\Delta \omega_n/\omega_0$ to be about 0.8 resulting in the corresponding frequency corrections $\varepsilon_4 \approx -0.48$ and $\varepsilon_3 \approx -0.28$ (see Fig.~\ref{fig:p-mode_sep_corr}).}, and discriminating it from other effects \citep[e.g. stellar density, radius, gravity, rotation, effective temperature and ionisation effects,][]{2009ApJ...700.1589S, 2009A&A...503L..21M, 2011MNRAS.418L.119H, 2020Natur.581..147B, 2021MNRAS.505.1476H} influencing $\Delta \omega_n/\omega_0$.

Furthermore, {by} substituting $\omega_n$ from Eq.~(\ref{eq:epsilon_n}) into Eq.~(\ref{eq:disp}) and treating $\varepsilon_n$ as a small parameter, we can obtain an approximate explicit analytical relationship between $\varepsilon_n$ and the acoustic cut-off frequency $\alpha$ as
\begin{equation}\label{eq:epsilon_approx}
\varepsilon_n \approx - \frac{n}{\sqrt{\alpha^2 - n^2\omega_0^2}},
\end{equation}
where both $\alpha$ and $\omega_0$ are normalised to the acoustic travel time $\tau_\mathrm{A}$. Our tests showed that the approximate solution Eq.~(\ref{eq:epsilon_approx}) agrees well with the full numerical solution of Eq.~(\ref{eq:disp}) for $|\varepsilon_n| \lesssim 0.1$, with the mismatch below 10\%.
\section{11-year frequency shifts}
\label{Sec:shifts}

In this section, we propose a mechanism for how the 11-year variability of the Sun's magnetic activity can influence p-mode oscillation frequencies \citep[see e.g.][]{1986Natur.323..603R, 1996BASI...24..199R} via modulation of the photospheric acoustic cut-off frequency, $\alpha$. Indeed, the effect of the magnetic field on $\alpha$ can be isolated as \citep[e.g.][]{2015A&A...582A..57A},
\begin{equation}\label{eq:cutoff_beta}
    \alpha = \frac{\alpha_\mathrm{HD}}{\sqrt{1+{\gamma\beta}/{2}}},
\end{equation}
where $\beta$ is the standard plasma parameter given by the ratio of the thermal pressure to the magnetic pressure, $\gamma$ is the standard adiabatic index, and $\alpha_\mathrm{HD}$ is the value of the acoustic cut-off fully determined by hydrodynamic parameters of the medium (i.e. sound speed and density scale height) in the infinite field regime ($\beta \to 0$).

\begin{figure*}
    \centering    
    \includegraphics[height=0.32\textwidth]{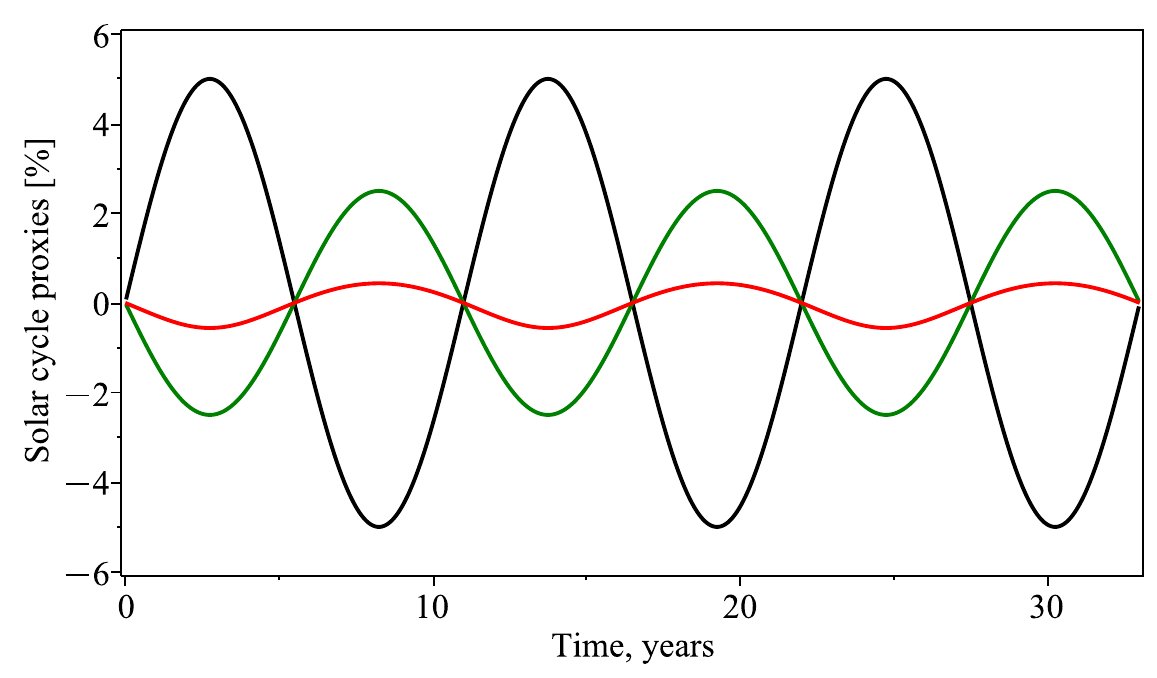}
    \includegraphics[height=0.32\textwidth]{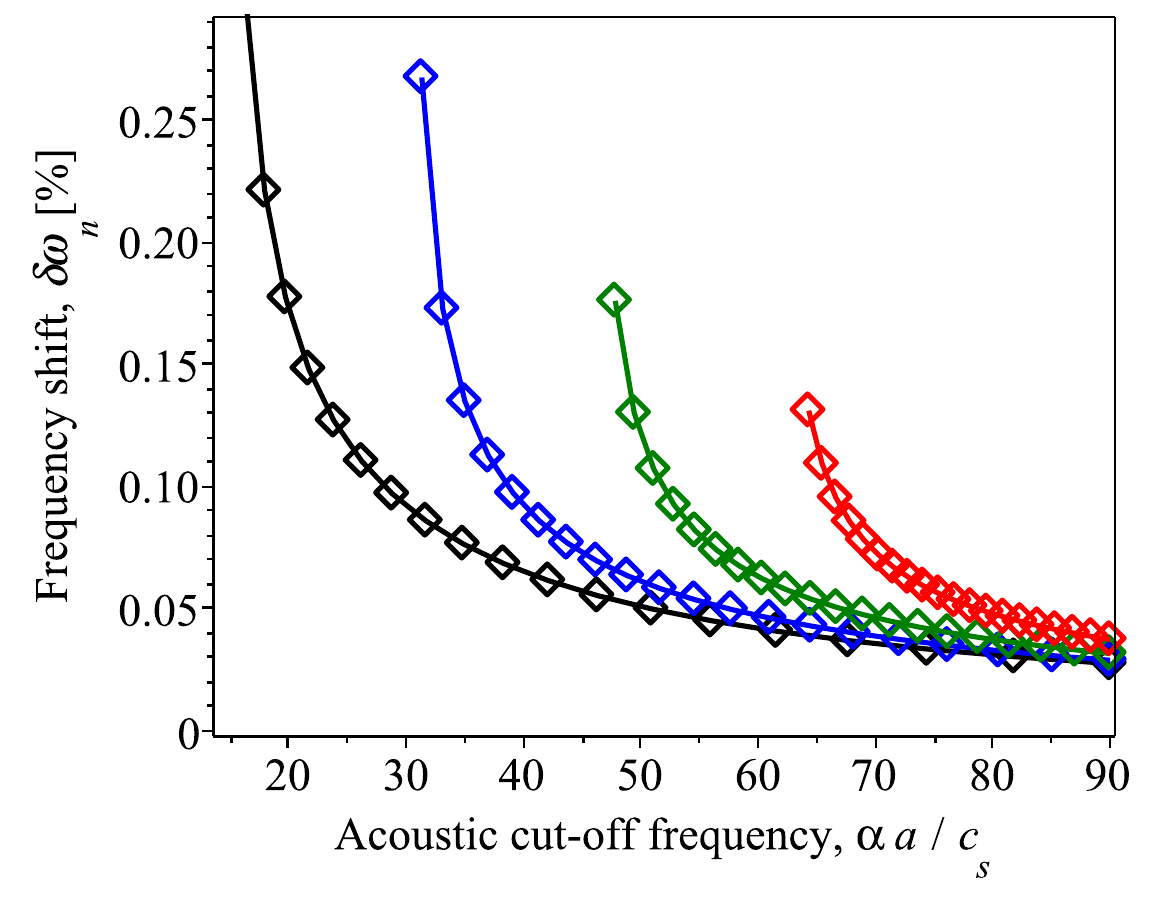}
    \caption{Left: Time profile of the 11-year periodic perturbation of the plasma parameter $\beta$ (ratio of the thermal pressure to the magnetic pressure) at the photosphere, $\delta\beta(t)$ (black) resulting in the perturbation of the acoustic cut-off frequency $(\alpha'-\alpha_0)/\alpha_0$ given by Eq.~(\ref{eq:cutoff_taylor}) (green) and in the perturbation of the p-mode frequency $[\omega_n(\alpha')-\omega_n(\alpha_0)]/\omega_n(\alpha_0)$ for $n=5$ ($\alpha_0\tau_\mathrm{A}=15$) (red), by the mechanism described in Sec.~\ref{Sec:shifts}.
    Right: Relative amplitude of the p-mode frequency shifts $\delta\omega_n = \left[\omega_n(\alpha_\mathrm{max})-\omega_n(\alpha_\mathrm{min})\right]/2\omega_n(\alpha_0)$ vs. acoustic cut-off frequency $\alpha_0$ for radial harmonics $n=5$ (black), 10 (blue), 15 (green), and 20 (red).}
    \label{fig:p-mode_freq_shift}
\end{figure*}

Demanding $\beta$ in Eq.~(\ref{eq:cutoff_beta}) to vary harmonically with the 11-year periodicity and small amplitude around some equilibrium value, $\beta = \beta_0 (1+\delta\beta)$, the perturbed acoustic cut-off $\alpha'$ can be Taylor-expanded around small parameter $\delta\beta$ as
\begin{equation}\label{eq:cutoff_taylor}
    \alpha' \approx \alpha_0\left(1-\frac{\delta\beta}{2}\right).
\end{equation}
Here, we used condition $\beta_0 \gg 1$ typical for the solar convection zone \citep[e.g.][]{2004LRSP....1....1F}, and $\alpha_0$ stands for the unperturbed value of the acoustic cut-off. Thus, taking a harmonic function with e.g. 5\% amplitude for $\delta\beta$ (see Fig.~\ref{fig:p-mode_freq_shift}, left panel), we obtain $\alpha'$ to vary in anti-phase with the same oscillation period. Such a behaviour of the acoustic cut-off with $\beta$ is generally consistent with observations of the 11-yr solar cycle, where the cut-off frequency is found to vary in phase with the radio flux \citep[e.g.][]{2011ApJ...743...99J}, which in turn has inverse proportionality to $\beta$.

\begin{figure}
    \centering    
    \includegraphics[width=\columnwidth]{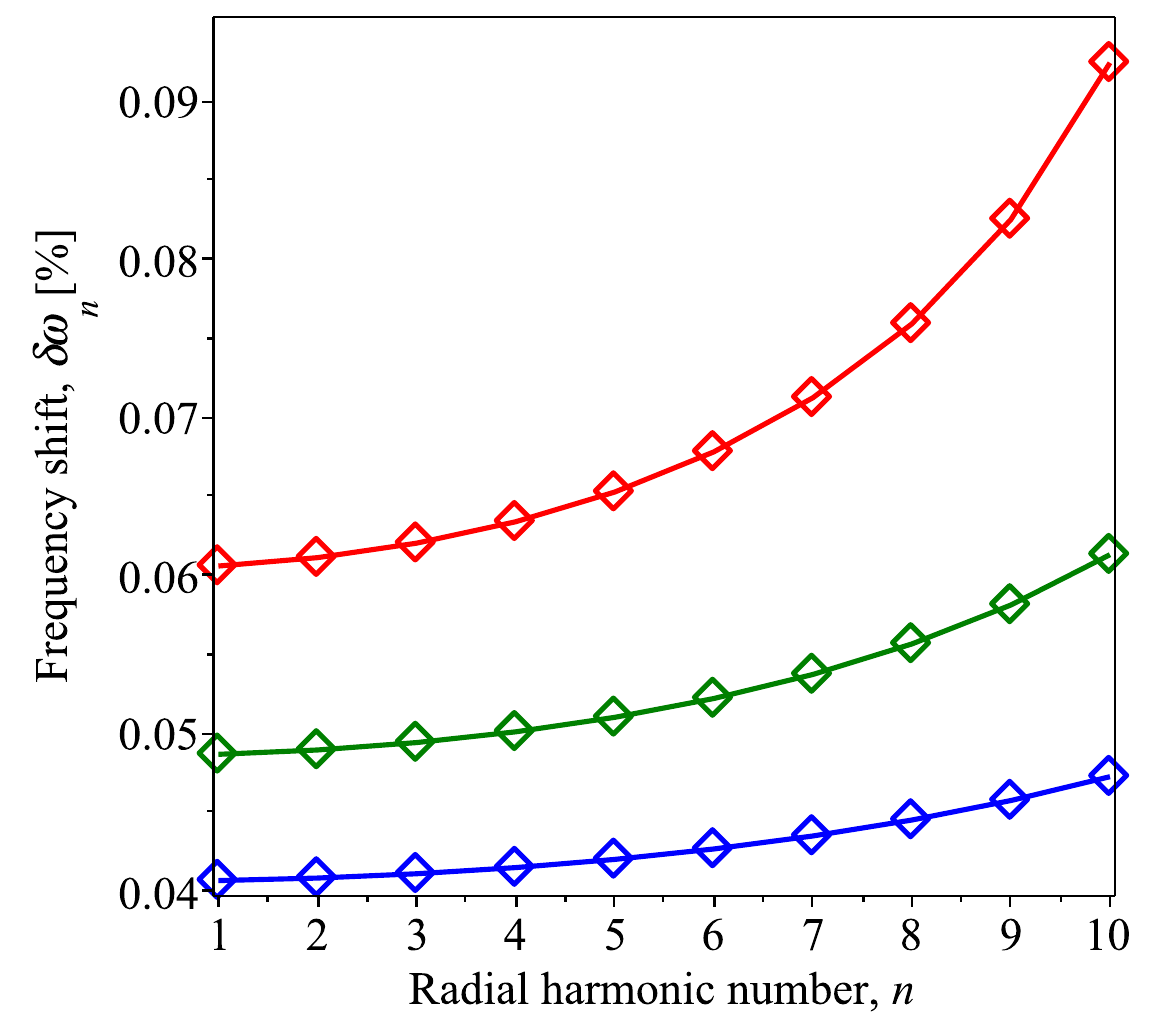}
    \caption{Dependence of the p-mode frequency shift amplitude $\delta\omega_n$ (see Fig.~\ref{fig:p-mode_freq_shift} and Sec.~\ref{Sec:shifts}) on the radial harmonic number $n$ for the acoustic cut-off frequency $\alpha_0\tau_\mathrm{A}=40$ (red), 50 (green), and 60 (blue).}
    \label{fig:p-mode_freq_shift_amp-n}
\end{figure}

We now substitute the acoustic cut-off $\alpha'$ Eq.~(\ref{eq:cutoff_taylor}) modulated by the 11-yr variability of the plasma parameter $\beta$ into our numerical solution of Eq.~(\ref{eq:disp}) for $\omega_n$ shown in Fig.~\ref{fig:p-mode_freq}, left panel. Thus, small-amplitude harmonic modulation of the cut-off frequency in the vicinity of $\alpha_0$ results in the corresponding small-amplitude periodic modulation of the p-mode oscillation frequency $\omega_n$, illustrated, for example, for the radial harmonic $n=5$ and $\alpha_0\tau_\mathrm{A}=15$ in the left-hand panel of Fig.~\ref{fig:p-mode_freq_shift}. The phase behaviour of $\omega_n$ modulated by this mechanism is prescribed, in general, by the sign of the gradient of the function $\omega_n(\alpha)$. In our case (Fig.~\ref{fig:p-mode_freq}), $\omega_n$ tends to increase with $\alpha$ (i.e. the gradient is positive), which results in the in-phase 11-year modulation of the acoustic cut-off and p-mode frequency shift (both are in anti-phase with the plasma parameter $\beta$) in Fig.~\ref{fig:p-mode_freq_shift}. The latter is also consistent with the observed behaviour \citep[e.g.][and references therein]{1985Natur.318..449W, 1989A&A...224..253P, 1990Natur.345..322E, 1990Natur.345..779L, 2014SSRv..186..191B}. In turn, the amplitude of the obtained frequency shift is prescribed by the value of the gradient of the function $\omega_n(\alpha)$. Thus, according to the left-hand panel in Fig.~\ref{fig:p-mode_freq_shift}, the arbitrarily chosen 5\% relative amplitude in the plasma parameter $\beta$ results in 2.5\% relative amplitude in the cut-off frequency $\alpha'$ ($\sim 125$\,$\mu$Hz for $\alpha_0 = 5000$\,$\mu$Hz) and in much lower relative amplitude in the p-mode frequency shift $\delta\omega_n$ as usually seen in observations. The right-hand panel in Fig.~\ref{fig:p-mode_freq_shift} shows the frequency shift amplitude $\delta\omega_n$ estimated as $\delta\omega_n = \left[\omega_n(\alpha_\mathrm{max})-\omega_n(\alpha_\mathrm{min})\right]/2\omega_n(\alpha_0)$ for radial harmonics $n=5$, 10, 15, and 20, which rapidly drops below 0.05\% with $\alpha$.
Furthermore, the obtained 11-yr frequency shift amplitudes $\delta\omega_n$ tend to increase with the radial harmonic number, $n$ (see Fig.~\ref{fig:p-mode_freq_shift_amp-n}). 
Thus, our mechanism readily reproduces the observed typical amplitudes of the p-mode frequency shift and its phase behaviour relative to other 11-yr solar cycle proxies.

\section{Discussion and Conclusions}
\label{sec:disc}

We studied the effects of the photospheric acoustic cut-off on the characteristic frequencies of p-modes trapped in a sub-photospheric acoustic cavity, using the Klein-Gordon equation with a piecewise acoustic potential. The developed approach allowed us to assess the corrections to the p-mode frequencies, caused by the cut-off effect, and to link the observed 11-yr helioseismic frequency shifts with the impact of the magnetic field on the photospheric acoustic cut-off frequency. In other words, by treating the sub-photospheric acoustic cavity as a resonator of a Fabry-P\'erot type and varying its reflection/transmission property (i.e. the photospheric acoustic cut-off frequency) periodically, we managed to reproduce the periodic frequency modulation of acoustic waves trapped inside the resonator. The main findings of this study can be summarised as follows:

\begin{itemize}
    \item Finite values of the photospheric acoustic cut-off frequency lead to the decrease in the characteristic frequencies of trapped p-modes. This is connected with the modification of the effective speed of acoustic waves by the cut-off effect. The revealed frequency correction is larger for higher radial harmonics, with frequencies closer to the acoustic cut-off frequency. Our estimations show that for $\ell=100$ (acoustic travel time through the cavity $\tau_\mathrm{A}\approx 20$\,min), the maximum decrease in the p-mode frequency can reach 200\,$\mu$Hz ({obtained as $\pi/2/2\pi/\tau_\mathrm{A}$}, for all radial harmonics).

    \item We derived an approximate but explicit relationship linking the expected frequency correction $\varepsilon_n$ with the acoustic cut-off frequency $\alpha$ and the radial harmonic number $n$ (Eq.~(\ref{eq:epsilon_approx})). Moreover, we proposed a scheme for estimating the frequency correction $\varepsilon_n$ in observations, using the large frequency separation $\Delta\omega_n$. In particular, the cut-off effect is shown to result in $\Delta\omega_n/\omega_0 < 1$, where $\omega_0 = \pi/\tau_\mathrm{A}$ (see Fig.~\ref{fig:p-mode_sep_corr}). This constraint can be useful for disentangling the cut-off effect from other effects influencing $\Delta\omega_n/\omega_0$ in future studies.
    
    \item The large frequency separation $\Delta\omega_n$ is found to decrease with the increasing radial harmonic number $n$ due to the cut-off effect. However, this decrease is found to be only mild, by 4--10\% for all cases considered.
    Observationally, the behaviour of $\Delta\omega_n$ with $n$ is seen to be non-monotonic for low-$\ell$ modes. It initially decreases with increasing $n$ at low $n$ and then increases for higher $n$ \citep[see e.g.][for the Sun and other stars]{2007MNRAS.381.1001K, 2013A&A...550A.126M, 2010A&A...524A..47G}, which may indicate the cut-off effect is counteracted by other mechanisms. In these low-$\ell$ observations, the cut-off frequency corresponds to around $n=36$ and we can readily see modes in a broad interval of $n$s (e.g. for $\ell=0$--2, we can observe $n=9$--30). For higher $\ell\simeq 100$, the cut-off is around $n=15$, and there $\Delta\omega_n$ does seem to decrease with increasing $n$ monotonically (see e.g. the standard SoHO/MDI frequency-degree diagram). The latter is broadly consistent with our result. However, a more detailed comparison, such as the decrease rate and {its} functional form predicted theoretically and seen in observations, would require a dedicated study.

    \item Low-amplitude periodic modulation of the plasma parameter $\beta$ at the photosphere is shown to result in the modulation of the acoustic cut-off and, more importantly, in the corresponding periodic shifts of trapped p-mode frequencies through the cut-off effect (see Fig.~\ref{fig:p-mode_freq_shift}). Using our model, we managed to reproduce the observed amplitudes of the 11-year frequency shift and its in-phase behaviour with other solar cycle proxies, such as radio flux (inversely proportional to $\beta$). This is in contrast to previous works discussed in Section \ref{sec:intro}, where a strong perturbation to the toroidal magnetic field required for reproducing the observed magnitude and frequency dependence of the p-mode frequency shifts. 

    \item  The detected dependence of the frequency shift amplitude on the radial harmonic number $n$ (Fig.~\ref{fig:p-mode_freq_shift_amp-n}) is very reminiscent of the trend observed in real data \citep[e.g.][]{1990Natur.345..779L, 2001MNRAS.324..910C, 2017SoPh..292...67B}, which is usually related to the frequency-dependence of the upper turning points of the modes and a near-surface perturbation of the modes by a magnetic field \citep{1980tsp..book.....C, 1990LNP...367..283G, 1991ApJ...370..752G}. Our model suggests that the cut-off effect may also play a role in those observations. A similar frequency (or $n$) dependence has also been observed on some other stars \citep[e.g.][]{2011A&A...530A.127S, 2016A&A...589A.118S}. However, other stars seem to exhibit an oscillatory relationship between frequency shift and frequency, which can potentially be attributed to a deeper seated magnetic field \citep[e.g.][]{2018A&A...611A..84S}. It would be interesting to determine whether the acoustic cut-off frequencies of these stars also vary with time.
\end{itemize}

The presented study is based on a 1D analytical model which has a number of important shortcomings. For example, we neglected the 2D effects responsible for both the acoustic wave refraction at the lower turning point and the possible direct impact of the magnetic Lorentz force on the modes. To account for these, a generalisation of our study using a more advanced 2D model of (magneto)acoustic-gravity waves in a stratified medium, described in e.g. \citet{2006RSPTA.364..447R, 2018MNRAS.480..623C}, is required. Likewise, our study can be expanded upon considering more realistic smooth radial profiles of the acoustic potential $V(r)$, thus incorporating the dependence of the upper turning point on the mode frequency. For example, \citet{2019ApJ...870...41O} used the WKB approximation to determine how the large frequency separation is affected by the acoustic cut-off effect. In their model, the acoustic cut-off is non-zero and varying in the solar interior, such that the mode frequencies are always greater than the acoustic cut-off. This means that the acoustic cut-off determines the radius of reflection and this radius varies from mode to mode. Our model of the acoustic cut-off as a step function is more simplistic, and was a necessary step in order to isolate the direct impact the acoustic cut-off has on mode frequencies, but adopting a set-up similar to that used bu \citet{2019ApJ...870...41O}, would represent and interesting extension to this work.  The exact analytical solution to the governing Klein-Gordon equation (\ref{eq:KG}) used in our study is derived under the assumption of a monochromatic harmonic driver, while in reality the convection zone motions driving p-mode oscillations are rather broadband and stochastic. Thus, the stochastic excitation of p-modes with the cut-off effect can also be studied in terms of our model by adjusting the source function on the right-hand side of Eq.~(\ref{eq:KG}). Finally, possible interactions of the evanescent part of p-modes, which was shown to reach up to the temperature minimum region in the Sun's atmosphere, with chromospheric resonators \citep[e.g.][]{2011ApJ...728...84B, 2014AstL...40..576Z} represent another interesting avenue for development. In all those studies, the low-dimensional results presented in this work can be used as a limiting case for comparison and validation.

\section*{Acknowledgements}

The work is supported by the STFC consolidated grant ST/X000915/1. DYK also acknowledge the Latvian Council of Science Project No. lzp2022/1-0017. The authors are grateful to Prof. Yvonne Elsworth FRS and Dr. Sergei Vorontsov for stimulating discussions.

\section*{Data Availability}
The data underlying this article are available in the article and in the references therein.



\bibliographystyle{mnras}

\begin{thebibliography}{}
	\makeatletter
	\relax
	\def\mn@urlcharsother{\let\do\@makeother \do\$\do\&\do\#\do\^\do\_\do\%\do\~}
	\def\mn@doi{\begingroup\mn@urlcharsother \@ifnextchar [ {\mn@doi@}
		{\mn@doi@[]}}
	\def\mn@doi@[#1]#2{\def\@tempa{#1}\ifx\@tempa\@empty \href
		{http://dx.doi.org/#2} {doi:#2}\else \href {http://dx.doi.org/#2} {#1}\fi
		\endgroup}
	\def\mn@eprint#1#2{\mn@eprint@#1:#2::\@nil}
	\def\mn@eprint@arXiv#1{\href {http://arxiv.org/abs/#1} {{\tt arXiv:#1}}}
	\def\mn@eprint@dblp#1{\href {http://dblp.uni-trier.de/rec/bibtex/#1.xml}
		{dblp:#1}}
	\def\mn@eprint@#1:#2:#3:#4\@nil{\def\@tempa {#1}\def\@tempb {#2}\def\@tempc
		{#3}\ifx \@tempc \@empty \let \@tempc \@tempb \let \@tempb \@tempa \fi \ifx
		\@tempb \@empty \def\@tempb {arXiv}\fi \@ifundefined
		{mn@eprint@\@tempb}{\@tempb:\@tempc}{\expandafter \expandafter \csname
			mn@eprint@\@tempb\endcsname \expandafter{\@tempc}}}
	
	\bibitem[\protect\citeauthoryear{{Aerts}, {Christensen-Dalsgaard}  \&
		{Kurtz}}{{Aerts} et~al.}{2010}]{2010aste.book.....A}
	{Aerts} C.,  {Christensen-Dalsgaard} J.,   {Kurtz} D.~W.,  2010,
	{Asteroseismology}, \mn@doi{10.1007/978-1-4020-5803-5.
	}
	
	\bibitem[\protect\citeauthoryear{{Afanasyev} \& {Nakariakov}}{{Afanasyev} \&
		{Nakariakov}}{2015}]{2015A&A...582A..57A}
	{Afanasyev} A.~N.,  {Nakariakov} V.~M.,  2015, \mn@doi [\aap]
	{10.1051/0004-6361/201526530}, \href
	{https://ui.adsabs.harvard.edu/abs/2015A&A...582A..57A} {582, A57}
	
	\bibitem[\protect\citeauthoryear{{Basu}}{{Basu}}{2016}]{2016LRSP...13....2B}
	{Basu} S.,  2016, \mn@doi [Living Reviews in Solar Physics]
	{10.1007/s41116-016-0003-4}, \href
	{https://ui.adsabs.harvard.edu/abs/2016LRSP...13....2B} {13, 2}
	
	\bibitem[\protect\citeauthoryear{{Basu}, {Broomhall}, {Chaplin}  \&
		{Elsworth}}{{Basu} et~al.}{2012}]{2012ApJ...758...43B}
	{Basu} S.,  {Broomhall} A.-M.,  {Chaplin} W.~J.,   {Elsworth} Y.,  2012,
	\mn@doi [\apj] {10.1088/0004-637X/758/1/43}, \href
	{https://ui.adsabs.harvard.edu/abs/2012ApJ...758...43B} {758, 43}
	
	\bibitem[\protect\citeauthoryear{{Bedding} et~al.,}{{Bedding}
		et~al.}{2020}]{2020Natur.581..147B}
	{Bedding} T.~R.,  et~al., 2020, \mn@doi [\nat] {10.1038/s41586-020-2226-8},
	\href {https://ui.adsabs.harvard.edu/abs/2020Natur.581..147B} {581, 147}
	
	\bibitem[\protect\citeauthoryear{{Botha}, {Arber}, {Nakariakov}  \&
		{Zhugzhda}}{{Botha} et~al.}{2011}]{2011ApJ...728...84B}
	{Botha} G.~J.~J.,  {Arber} T.~D.,  {Nakariakov} V.~M.,   {Zhugzhda} Y.~D.,
	2011, \mn@doi [\apj] {10.1088/0004-637X/728/2/84}, \href
	{https://ui.adsabs.harvard.edu/abs/2011ApJ...728...84B} {728, 84}
	
	\bibitem[\protect\citeauthoryear{{Broomhall}}{{Broomhall}}{2017}]{2017SoPh..292...67B}
	{Broomhall} A.~M.,  2017, \mn@doi [\solphys] {10.1007/s11207-017-1068-5}, \href
	{https://ui.adsabs.harvard.edu/abs/2017SoPh..292...67B} {292, 67}
	
	\bibitem[\protect\citeauthoryear{{Broomhall}, {Chatterjee}, {Howe}, {Norton}
		\& {Thompson}}{{Broomhall} et~al.}{2014}]{2014SSRv..186..191B}
	{Broomhall} A.~M.,  {Chatterjee} P.,  {Howe} R.,  {Norton} A.~A.,   {Thompson}
	M.~J.,  2014, \mn@doi [\ssr] {10.1007/s11214-014-0101-3}, \href
	{https://ui.adsabs.harvard.edu/abs/2014SSRv..186..191B} {186, 191}
	
	\bibitem[\protect\citeauthoryear{{Chaplin}, {Appourchaux}, {Elsworth}, {Isaak}
		\& {New}}{{Chaplin} et~al.}{2001}]{2001MNRAS.324..910C}
	{Chaplin} W.~J.,  {Appourchaux} T.,  {Elsworth} Y.,  {Isaak} G.~R.,   {New} R.,
	2001, \mn@doi [\mnras] {10.1046/j.1365-8711.2001.04357.x}, \href
	{https://ui.adsabs.harvard.edu/abs/2001MNRAS.324..910C} {324, 910}
	
	\bibitem[\protect\citeauthoryear{{Christensen-Dalsgaard}}{{Christensen-Dalsgaard}}{2021}]{2021LRSP...18....2C}
	{Christensen-Dalsgaard} J.,  2021, \mn@doi [Living Reviews in Solar Physics]
	{10.1007/s41116-020-00028-3}, \href
	{https://ui.adsabs.harvard.edu/abs/2021LRSP...18....2C} {18, 2}
	
	\bibitem[\protect\citeauthoryear{{Costa}, {Schneiter}  \& {Zurbriggen}}{{Costa}
		et~al.}{2018}]{2018MNRAS.480..623C}
	{Costa} A.,  {Schneiter} M.,   {Zurbriggen} E.,  2018, \mn@doi [\mnras]
	{10.1093/mnras/sty1828}, \href
	{https://ui.adsabs.harvard.edu/abs/2018MNRAS.480..623C} {480, 623}
	
	\bibitem[\protect\citeauthoryear{{Cox}}{{Cox}}{1980}]{1980tsp..book.....C}
	{Cox} J.~P.,  1980, {Theory of Stellar Pulsation. (PSA-2), Volume 2}.
	Vol. 2
	
	\bibitem[\protect\citeauthoryear{{Deubner} \& {Gough}}{{Deubner} \&
		{Gough}}{1984}]{1984ARA&A..22..593D}
	{Deubner} F.-L.,  {Gough} D.,  1984, \mn@doi [\araa]
	{10.1146/annurev.aa.22.090184.003113}, \href
	{https://ui.adsabs.harvard.edu/abs/1984ARA&A..22..593D} {22, 593}
	
	\bibitem[\protect\citeauthoryear{{Duvall}}{{Duvall}}{1982}]{1982Natur.300..242D}
	{Duvall} T.~L. J.,  1982, \mn@doi [\nat] {10.1038/300242a0}, \href
	{https://ui.adsabs.harvard.edu/abs/1982Natur.300..242D} {300, 242}
	
	\bibitem[\protect\citeauthoryear{{Dziembowski} \& {Goode}}{{Dziembowski} \&
		{Goode}}{2005}]{2005ApJ...625..548D}
	{Dziembowski} W.~A.,  {Goode} P.~R.,  2005, \mn@doi [\apj] {10.1086/429712},
	\href {https://ui.adsabs.harvard.edu/abs/2005ApJ...625..548D} {625, 548}
	
	\bibitem[\protect\citeauthoryear{{Elsworth}, {Howe}, {Isaak}, {McLeod}  \&
		{New}}{{Elsworth} et~al.}{1990}]{1990Natur.345..322E}
	{Elsworth} Y.,  {Howe} R.,  {Isaak} G.~R.,  {McLeod} C.~P.,   {New} R.,  1990,
	\mn@doi [\nat] {10.1038/345322a0}, \href
	{https://ui.adsabs.harvard.edu/abs/1990Natur.345..322E} {345, 322}
	
	\bibitem[\protect\citeauthoryear{{Fan}}{{Fan}}{2004}]{2004LRSP....1....1F}
	{Fan} Y.,  2004, \mn@doi [Living Reviews in Solar Physics]
	{10.12942/lrsp-2004-1}, \href
	{https://ui.adsabs.harvard.edu/abs/2004LRSP....1....1F} {1, 1}
	
	\bibitem[\protect\citeauthoryear{{Foullon} \& {Roberts}}{{Foullon} \&
		{Roberts}}{2005}]{2005A&A...439..713F}
	{Foullon} C.,  {Roberts} B.,  2005, \mn@doi [\aap]
	{10.1051/0004-6361:20041910}, \href
	{https://ui.adsabs.harvard.edu/abs/2005A&A...439..713F} {439, 713}
	
	\bibitem[\protect\citeauthoryear{{Fr{\"o}hlich} et~al.,}{{Fr{\"o}hlich}
		et~al.}{1997}]{1997SoPh..170....1F}
	{Fr{\"o}hlich} C.,  et~al., 1997, \mn@doi [\solphys] {10.1023/A:1004969622753},
	\href {https://ui.adsabs.harvard.edu/abs/1997SoPh..170....1F} {170, 1}
	
	\bibitem[\protect\citeauthoryear{{Garc{\'\i}a} \& {Ballot}}{{Garc{\'\i}a} \&
		{Ballot}}{2019}]{2019LRSP...16....4G}
	{Garc{\'\i}a} R.~A.,  {Ballot} J.,  2019, \mn@doi [Living Reviews in Solar
	Physics] {10.1007/s41116-019-0020-1}, \href
	{https://ui.adsabs.harvard.edu/abs/2019LRSP...16....4G} {16, 4}
	
	\bibitem[\protect\citeauthoryear{{Gaulme} et~al.,}{{Gaulme}
		et~al.}{2010}]{2010A&A...524A..47G}
	{Gaulme} P.,  et~al., 2010, \mn@doi [\aap] {10.1051/0004-6361/201014142}, \href
	{https://ui.adsabs.harvard.edu/abs/2010A&A...524A..47G} {524, A47}
	
	\bibitem[\protect\citeauthoryear{{Goldreich}, {Murray}, {Willette}  \&
		{Kumar}}{{Goldreich} et~al.}{1991}]{1991ApJ...370..752G}
	{Goldreich} P.,  {Murray} N.,  {Willette} G.,   {Kumar} P.,  1991, \mn@doi
	[\apj] {10.1086/169858}, \href
	{https://ui.adsabs.harvard.edu/abs/1991ApJ...370..752G} {370, 752}
	
	\bibitem[\protect\citeauthoryear{{Gough}}{{Gough}}{1990}]{1990LNP...367..283G}
	{Gough} D.~O.,  1990, in {Osaki} Y.,  {Shibahashi} H.,  eds, , Vol.~367,
	Progress of Seismology of the Sun and Stars.
	p.~283, \mn@doi{10.1007/3-540-53091-610.1007/3-540-53091-6_93}
	
	\bibitem[\protect\citeauthoryear{{Gough}}{{Gough}}{1993}]{1993afd..conf..399G}
	{Gough} D.~O.,  1993, in Astrophysical Fluid Dynamics - Les Houches 1987. pp
	399--560
	
	\bibitem[\protect\citeauthoryear{{Hasanzadeh}, {Safari}  \&
		{Ghasemi}}{{Hasanzadeh} et~al.}{2021}]{2021MNRAS.505.1476H}
	{Hasanzadeh} A.,  {Safari} H.,   {Ghasemi} H.,  2021, \mn@doi [\mnras]
	{10.1093/mnras/stab1411}, \href
	{https://ui.adsabs.harvard.edu/abs/2021MNRAS.505.1476H} {505, 1476}
	
	\bibitem[\protect\citeauthoryear{{Hekker}, {Basu}, {Elsworth}  \&
		{Chaplin}}{{Hekker} et~al.}{2011}]{2011MNRAS.418L.119H}
	{Hekker} S.,  {Basu} S.,  {Elsworth} Y.,   {Chaplin} W.~J.,  2011, \mn@doi
	[\mnras] {10.1111/j.1745-3933.2011.01156.x}, \href
	{https://ui.adsabs.harvard.edu/abs/2011MNRAS.418L.119H} {418, L119}
	
	\bibitem[\protect\citeauthoryear{{Jim{\'e}nez}, {Jim{\'e}nez-Reyes}  \&
		{Garc{\'\i}a}}{{Jim{\'e}nez} et~al.}{2005}]{2005ApJ...623.1215J}
	{Jim{\'e}nez} A.,  {Jim{\'e}nez-Reyes} S.~J.,   {Garc{\'\i}a} R.~A.,  2005,
	\mn@doi [\apj] {10.1086/428879}, \href
	{https://ui.adsabs.harvard.edu/abs/2005ApJ...623.1215J} {623, 1215}
	
	\bibitem[\protect\citeauthoryear{{Jim{\'e}nez}, {Garc{\'\i}a}  \&
		{Pall{\'e}}}{{Jim{\'e}nez} et~al.}{2011}]{2011ApJ...743...99J}
	{Jim{\'e}nez} A.,  {Garc{\'\i}a} R.~A.,   {Pall{\'e}} P.~L.,  2011, \mn@doi
	[\apj] {10.1088/0004-637X/743/2/99}, \href
	{https://ui.adsabs.harvard.edu/abs/2011ApJ...743...99J} {743, 99}
	
	\bibitem[\protect\citeauthoryear{{Karoff}}{{Karoff}}{2007}]{2007MNRAS.381.1001K}
	{Karoff} C.,  2007, \mn@doi [\mnras] {10.1111/j.1365-2966.2007.12340.x}, \href
	{https://ui.adsabs.harvard.edu/abs/2007MNRAS.381.1001K} {381, 1001}
	
	\bibitem[\protect\citeauthoryear{{Kiefer} \& {Roth}}{{Kiefer} \&
		{Roth}}{2018}]{2018ApJ...854...74K}
	{Kiefer} R.,  {Roth} M.,  2018, \mn@doi [\apj] {10.3847/1538-4357/aaa3f7},
	\href {https://ui.adsabs.harvard.edu/abs/2018ApJ...854...74K} {854, 74}
	
	\bibitem[\protect\citeauthoryear{{Kosak}, {Kiefer}  \& {Broomhall}}{{Kosak}
		et~al.}{2022}]{2022MNRAS.512.5743K}
	{Kosak} K.,  {Kiefer} R.,   {Broomhall} A.~M.,  2022, \mn@doi [\mnras]
	{10.1093/mnras/stac647}, \href
	{https://ui.adsabs.harvard.edu/abs/2022MNRAS.512.5743K} {512, 5743}
	
	\bibitem[\protect\citeauthoryear{{Kumar} \& {Lu}}{{Kumar} \&
		{Lu}}{1991}]{1991ApJ...375L..35K}
	{Kumar} P.,  {Lu} E.,  1991, \mn@doi [\apjl] {10.1086/186082}, \href
	{https://ui.adsabs.harvard.edu/abs/1991ApJ...375L..35K} {375, L35}
	
	\bibitem[\protect\citeauthoryear{{Libbrecht} \& {Woodard}}{{Libbrecht} \&
		{Woodard}}{1990}]{1990Natur.345..779L}
	{Libbrecht} K.~G.,  {Woodard} M.~F.,  1990, \mn@doi [\nat] {10.1038/345779a0},
	\href {https://ui.adsabs.harvard.edu/abs/1990Natur.345..779L} {345, 779}
	
	\bibitem[\protect\citeauthoryear{{Miglio} et~al.,}{{Miglio}
		et~al.}{2009}]{2009A&A...503L..21M}
	{Miglio} A.,  et~al., 2009, \mn@doi [\aap] {10.1051/0004-6361/200912822}, \href
	{https://ui.adsabs.harvard.edu/abs/2009A&A...503L..21M} {503, L21}
	
	\bibitem[\protect\citeauthoryear{{Mosser} et~al.,}{{Mosser}
		et~al.}{2013}]{2013A&A...550A.126M}
	{Mosser} B.,  et~al., 2013, \mn@doi [\aap] {10.1051/0004-6361/201220435}, \href
	{https://ui.adsabs.harvard.edu/abs/2013A&A...550A.126M} {550, A126}
	
	\bibitem[\protect\citeauthoryear{{Nigam} \& {Kosovichev}}{{Nigam} \&
		{Kosovichev}}{1998}]{1998ApJ...505L..51N}
	{Nigam} R.,  {Kosovichev} A.~G.,  1998, \mn@doi [\apjl] {10.1086/311594}, \href
	{https://ui.adsabs.harvard.edu/abs/1998ApJ...505L..51N} {505, L51}
	
	\bibitem[\protect\citeauthoryear{{Ong} \& {Basu}}{{Ong} \&
		{Basu}}{2019}]{2019ApJ...870...41O}
	{Ong} J.~M.~J.,  {Basu} S.,  2019, \mn@doi [\apj] {10.3847/1538-4357/aaf1b5},
	\href {https://ui.adsabs.harvard.edu/abs/2019ApJ...870...41O} {870, 41}
	
	\bibitem[\protect\citeauthoryear{{Palle}, {Regulo}  \& {Roca Cortes}}{{Palle}
		et~al.}{1989}]{1989A&A...224..253P}
	{Palle} P.~L.,  {Regulo} C.,   {Roca Cortes} T.,  1989, \aap, \href
	{https://ui.adsabs.harvard.edu/abs/1989A&A...224..253P} {224, 253}
	
	\bibitem[\protect\citeauthoryear{{Roberts}}{{Roberts}}{1996}]{1996BASI...24..199R}
	{Roberts} B.,  1996, Bulletin of the Astronomical Society of India, \href
	{https://ui.adsabs.harvard.edu/abs/1996BASI...24..199R} {24, 199}
	
	\bibitem[\protect\citeauthoryear{{Roberts}}{{Roberts}}{2006}]{2006RSPTA.364..447R}
	{Roberts} B.,  2006, \mn@doi [Philosophical Transactions of the Royal Society
	of London Series A] {10.1098/rsta.2005.1709}, \href
	{https://ui.adsabs.harvard.edu/abs/2006RSPTA.364..447R} {364, 447}
	
	\bibitem[\protect\citeauthoryear{{Roberts} \& {Campbell}}{{Roberts} \&
		{Campbell}}{1986}]{1986Natur.323..603R}
	{Roberts} B.,  {Campbell} W.~R.,  1986, \mn@doi [\nat] {10.1038/323603a0},
	\href {https://ui.adsabs.harvard.edu/abs/1986Natur.323..603R} {323, 603}
	
	\bibitem[\protect\citeauthoryear{{Roxburgh} \& {Vorontsov}}{{Roxburgh} \&
		{Vorontsov}}{1995}]{1995MNRAS.272..850R}
	{Roxburgh} I.~W.,  {Vorontsov} S.~V.,  1995, \mn@doi [\mnras]
	{10.1093/mnras/272.4.850}, \href
	{https://ui.adsabs.harvard.edu/abs/1995MNRAS.272..850R} {272, 850}
	
	\bibitem[\protect\citeauthoryear{{Salabert}, {R{\'e}gulo}, {Ballot},
		{Garc{\'\i}a}  \& {Mathur}}{{Salabert} et~al.}{2011}]{2011A&A...530A.127S}
	{Salabert} D.,  {R{\'e}gulo} C.,  {Ballot} J.,  {Garc{\'\i}a} R.~A.,   {Mathur}
	S.,  2011, \mn@doi [\aap] {10.1051/0004-6361/201116633}, \href
	{https://ui.adsabs.harvard.edu/abs/2011A&A...530A.127S} {530, A127}
	
	\bibitem[\protect\citeauthoryear{{Salabert} et~al.,}{{Salabert}
		et~al.}{2016}]{2016A&A...589A.118S}
	{Salabert} D.,  et~al., 2016, \mn@doi [\aap] {10.1051/0004-6361/201527978},
	\href {https://ui.adsabs.harvard.edu/abs/2016A&A...589A.118S} {589, A118}
	
	\bibitem[\protect\citeauthoryear{{Salabert}, {R{\'e}gulo}, {P{\'e}rez
			Hern{\'a}ndez}  \& {Garc{\'\i}a}}{{Salabert}
		et~al.}{2018}]{2018A&A...611A..84S}
	{Salabert} D.,  {R{\'e}gulo} C.,  {P{\'e}rez Hern{\'a}ndez} F.,   {Garc{\'\i}a}
	R.~A.,  2018, \mn@doi [\aap] {10.1051/0004-6361/201731714}, \href
	{https://ui.adsabs.harvard.edu/abs/2018A&A...611A..84S} {611, A84}
	
	\bibitem[\protect\citeauthoryear{{Stello} et~al.,}{{Stello}
		et~al.}{2009}]{2009ApJ...700.1589S}
	{Stello} D.,  et~al., 2009, \mn@doi [\apj] {10.1088/0004-637X/700/2/1589},
	\href {https://ui.adsabs.harvard.edu/abs/2009ApJ...700.1589S} {700, 1589}
	
	\bibitem[\protect\citeauthoryear{{Taroyan} \& {Erd{\'e}lyi}}{{Taroyan} \&
		{Erd{\'e}lyi}}{2008}]{2008SoPh..251..523T}
	{Taroyan} Y.,  {Erd{\'e}lyi} R.,  2008, \mn@doi [\solphys]
	{10.1007/s11207-008-9154-3}, \href
	{https://ui.adsabs.harvard.edu/abs/2008SoPh..251..523T} {251, 523}
	
	\bibitem[\protect\citeauthoryear{{Vorontsov}, {Jefferies}, {Duval}  \&
		{Harvey}}{{Vorontsov} et~al.}{1998}]{1998MNRAS.298..464V}
	{Vorontsov} S.~V.,  {Jefferies} S.~M.,  {Duval} T.~L. J.,   {Harvey} J.~W.,
	1998, \mn@doi [\mnras] {10.1046/j.1365-8711.1998.01630.x}, \href
	{https://ui.adsabs.harvard.edu/abs/1998MNRAS.298..464V} {298, 464}
	
	\bibitem[\protect\citeauthoryear{{Woodard} \& {Noyes}}{{Woodard} \&
		{Noyes}}{1985}]{1985Natur.318..449W}
	{Woodard} M.~F.,  {Noyes} R.~W.,  1985, \mn@doi [\nat] {10.1038/318449a0},
	\href {https://ui.adsabs.harvard.edu/abs/1985Natur.318..449W} {318, 449}
	
	\bibitem[\protect\citeauthoryear{{Zhugzhda} \& {Sych}}{{Zhugzhda} \&
		{Sych}}{2014}]{2014AstL...40..576Z}
	{Zhugzhda} Y.~D.,  {Sych} R.~A.,  2014, \mn@doi [Astronomy Letters]
	{10.1134/S1063773714090059}, \href
	{https://ui.adsabs.harvard.edu/abs/2014AstL...40..576Z} {40, 576}
	
	\makeatother
\end{thebibliography}




%
%
%

\bsp	
\label{lastpage}
\end{document}